\documentclass[fleqn,usenatbib]{mnras}

\usepackage{newtxtext,newtxmath}

\usepackage[T1]{fontenc}

\DeclareRobustCommand{\VAN}[3]{#2}
\let\VANthebibliography\thebibliography
\def\thebibliography{\DeclareRobustCommand{\VAN}[3]{##3}\VANthebibliography}

\usepackage{graphicx}	
\usepackage{amsmath}	
\usepackage{subcaption}
\usepackage{lineno}

\newcommand{\hi}{H$\rm_I$}

\newcommand{\hiNoHiImprovementMS}{99\%}

\newcommand{\hiGwNoHiGwImprovementM}{34\%}
\newcommand{\hiGwNoHiGwImprovementS}{47\%}
\newcommand{\hiGwPrecisionMS}{20\%}

\newcommand{\allNoHiImprovementM}{61\%}
\newcommand{\allNoHiImprovementS}{53\%}

\newcommand{\allGwPrecisionM}{10\%}
\newcommand{\allGwPrecisionS}{15\%}

\usepackage{orcidlink}
\usepackage{here}

\title[Dark Sirens with 21-cm Intensity Mapping]{Cosmological Inference with Gravitational-Wave Dark Sirens and 21-cm Intensity Mapping: Forecasts for LIGO-Virgo-KAGRA with MeerKAT and SKAO}

\author[D. Nanadoumgar-Lacroze et al.]{Dounia Nanadoumgar-Lacroze$^{1}$\thanks{E-mail: dnala@ifae.es (DNL)}\thanks{This work is part of the doctoral thesis of DNL within the framework of the Doctoral Program in Physics at the Autonomous University of Barcelona.}\orcidlink{0009-0009-7255-8111},
Konstantin Leyde$^{2,3}$\thanks{E-mail: kleyde@flatironinstitute.org (KL)}\orcidlink{0000-0001-7661-2810},
Steven Cunnington$^{3}$\thanks{E-mail: steve.cunnington@port.ac.uk (SC)}\orcidlink{0000-0001-6594-107X},
Tessa Baker$^{3}$\orcidlink{0000-0001-5470-7616},
\newauthor
Ian Harrison$^{4}$\orcidlink{0000-0002-4437-0770},
Ulyana Dupletsa$^{5}$\orcidlink{0000-0003-2766-247X},
Simone Mastrogiovanni$^{6}$\orcidlink{0000-0003-1606-4183},
Riccardo Murgia$^{7,8,9}$\orcidlink{0000-0002-2224-7704},
\newauthor
Tommaso Ronconi$^{10}$\orcidlink{0000-0002-3515-6801},
Matteo Schulz$^{11,12}$\orcidlink{0009-0005-8184-0232},
Marta Spinelli$^{13,14}$\orcidlink{0000-0003-0148-3254}
\\
$^{1}$Institut de Física d’Altes Energies, The Barcelona Institute of Science and Technology, Campus UAB, E-08193 Bellaterra, Barcelona, Spain\\
$^{2}$Center for Computational Astrophysics, Flatiron Institute, 162, 5th Avenue, New York, NY 10010, USA\\
$^{3}$Institute of Cosmology \& Gravitation, University of Portsmouth, Dennis Sciama Building, Portsmouth, PO1 3FX, UK\\
$^{4}$School of Physics and Astronomy, Cardiff University, The Parade, Cardiff, Wales CF24 3AA, UK\\
$^{5}$Marietta Blau Institute - Austrian Academy of Sciences, 1010 Vienna, Austria\\
$^{6}$INFN, Sezione di Roma, 1-00185 Roma, Italy\\
$^{7}$Dipartimento di Fisica, Università degli Studi di Cagliari, Cittadella Universitaria, 09042 Monserrato (CA), Italy\\
$^{8}$INFN, Sezione di Cagliari, Cittadella Universitaria, 09042 Monserrato (CA), Italy\\
$^{9}$INAF - Osservatorio Astronomico di Cagliari, Via della Scienza 5, 09047 Selargius (CA), Italy\\
$^{10}$INAF - Institute of Radioastronomy (IRA), Via Gobetti 101, 40129 Bologna, Italy\\
$^{11}$Gran Sasso Science Institute (GSSI), Viale F. Crispi 7, L’Aquila (AQ), I-67100, Italy\\
$^{12}$INFN - Laboratori Nazionali del Gran Sasso (LNGS), L’Aquila (AQ), I-67100, Italy\\
$^{13}$Observatoire de la Côte d’Azur, Laboratoire Lagrange, Bd de l’Observatoire, 06304 Nice, France\\
$^{14}$Department of Physics \& Astronomy, University of the Western Cape, Cape Town 7535, South Africa\\
}

\date{Accepted XXX. Received YYY; in original form ZZZ}

\pubyear{\the\year{}}

\begin{document}
\label{firstpage}
\pagerange{\pageref{firstpage}--\pageref{lastpage}}
\maketitle

\begin{abstract}
Gravitational waves and 21-cm intensity mapping are emerging as complementary cosmological probes: gravitational waves provide direct luminosity-distance measurements, while neutral-hydrogen intensity mapping traces the matter distribution in redshift space, producing large-scale three-dimensional maps of the 21-cm signal. Their synergy offers a promising avenue for precise measurements of the Hubble constant, $H_0$. Within a standard-siren framework, we perform a joint cosmological and population analysis using $\mathcal{O}(150)$ mock gravitational-wave detections with signal-to-noise ratios above 12, as projected for one year of observations at LIGO--Virgo--KAGRA Observing Run~5 sensitivity. We combine these with mock 21-cm intensity maps from MeerKLASS (UHF band) and SKA-Mid (Band~1), and study how observational and instrumental limitations in both the gravitational-wave and intensity-mapping data affect the $H_0$ constraints. We find that uncertainty in gravitational-wave parameter measurements is a dominant limitation on the cosmological constraints. When this uncertainty is neglected, the inclusion of 21-cm information enables sub-percent constraints on $H_0$. We also find that observational effects in the 21-cm surveys substantially weaken the constraints, although they remain tighter than those obtained without 21-cm information (spectral-siren analysis). Despite accounting for beam convolution, instrumental noise, partial sky coverage, and foreground cleaning, including 21-cm information reduces the $H_0$ uncertainty relative to the spectral-siren results by \allNoHiImprovementM{} for MeerKLASS and \allNoHiImprovementS{} for SKA-Mid. Because our results are based on a single simulation realisation, they should be interpreted in light of possible realisation-to-realisation variability. Beam convolution most severely weakens the constraints. Instrumental noise can induce multimodality in the posterior, potentially biasing the inferred parameters. This effect is particularly pronounced for MeerKLASS, which has a higher noise level. The results are sensitive to foreground-cleaning choices, with both under-cleaning and over-cleaning degrading the accuracy. Partial sky coverage has a relatively small impact. In the most realistic scenario, which accounts for both gravitational-wave parameter-estimation uncertainties and intensity-mapping observational effects, we obtain $H_0$ precision estimates of \allGwPrecisionM{} and \allGwPrecisionS{} for MeerKLASS and SKA-Mid, respectively.
\end{abstract}

\begin{keywords}
gravitational waves -- radio lines: general -- cosmological parameters
\end{keywords}



\section{Introduction}
Gravitational-wave (GW) detections from compact binary coalescences (CBCs) provide a direct measurement of the luminosity distance, enabling their use as \textit{standard sirens} when associated with an independent redshift measurement~\citep{Schutz_1986}. Establishing a direct relationship between redshift and luminosity distance is key to resolving the Hubble tension~\citep{Planck2020,Riess2022}. GW cosmology offers a new, independent way to constrain the present-day expansion history of the universe. We distinguish between \textit{bright sirens} when an electromagnetic counterpart is identified, and \textit{dark sirens} otherwise. For instance, the gamma-ray counterpart to the binary neutron star GW170817~\citep{GW170817_detection, GW170817_MM} enabled a precise redshift estimation, leading to a tight constraint on $H_0$~\citep{GW170817_H0}. However, most events detected by the LIGO-Virgo-KAGRA (LVK) network are binary black hole (BBH) mergers~\citep{GWTC4_cat} which motivates the development of alternative methods to infer redshift. A possible approach is to construct a redshift prior from the distribution of matter in the sky region of the GW event. This method, known as the \textit{galaxy-catalogue method}, was first introduced using galaxy surveys~\citep{Schutz_1986,Del_Pozzo_2012,Fishbach_2019,Soares-Santos_2019,Palmese_2020,Gray_2020,Gray_2022}, as compact objects are expected to form and merge in dense environments with a high star-formation rate (SFR) such as galaxies. This approach is used in combination with the \textit{spectral sirens} method, leveraging features in the redshifted mass distribution of the GW population, to jointly infer cosmological and population parameters~\citep{Gray_2023,Mastrogiovanni_2023,GWTC5_cosmo}. The latest LVK results constrain $H_0$ to approximately 20\% using dark sirens alone, improving to approximately 12\% when the bright siren GW170817 is included \citep{GWTC5_cosmo}. This is, however, not yet sufficient to address the Hubble tension. The constraining power of the galaxy-catalogue method is limited by catalogue incompleteness across relevant redshifts and by the difficulty of accurately characterising completeness as a function of redshift and galaxy properties. Without precise galaxy redshift estimates and sufficient galaxy support in the GW events three-dimensional localisation volume, the efficiency of the galaxy-catalogue method is significantly reduced~\citep{Gray_2020, GWTC5_cosmo}. Heterogeneous sky coverage across combined galaxy surveys further complicates the modelling of catalogue completeness. Finally, large GW distance and sky-localisation uncertainties result in broad three-dimensional localisation volumes containing many potential host galaxies, limiting the constraining power of the analysis.

This work considers neutral hydrogen (\hi{}) intensity mapping (IM) \citep{Bharadwaj:2000av,Battye:2004re,Wyithe:2007rq,Chang:2007xk,Kovetz_2017,Liu:2019awk,Bernal_2022} which allows for the rapid surveying of large cosmological volumes by charting emission from the 21-cm hyperfine transition line, observed at radio frequencies ($\nu_0=1420.405~\rm{MHz}$). The abundance of \hi \ and its association with large-scale over-densities \citep{Villaescusa-Navarro_2018} make it an indirect indicator of the regions where CBCs are most likely to occur. It provides an alternative to optical electromagnetic tracers that can mitigate the limited completeness of current spectroscopic galaxy surveys at $z \gtrsim 0.5$, while offering much finer redshift resolution than is typically achievable with pre-stage~IV optical surveys. Single-dish IM observations, such as those from the SKA Observatory (SKAO) \citep{SKA_Cosmology_SWG_2020} and MeerKAT's Large Area Synoptic Survey (MeerKLASS) \citep{Santos_2017}, have a poorer angular resolution, observationally limited by the radio telescope beam size. Both ongoing and near-future IM surveys are expected to reach $\gtrsim 10{,}000\,$deg$^2$ of sky coverage, with e.g. MeerKLASS \citep{Cunnington:2025sdr} making them a particularly promising resource for advancing GW cosmology.

The synergy between GW sources and \hi\ IM has been previously discussed by \cite{Scelfo_2022} and \cite{Schulz_2026} to probe astrophysical and cosmological models with cross-correlation methods, and by \cite{Mukherjee_2022} to study CBC formation channels. Instead, we forecast how \hi\ IM support enhances cosmological constraints from dark sirens by extending redshift support beyond the galaxy catalogue regime. An analogous approach, referred to as \textit{radio sirens}, was investigated in \cite{Dupletsa_2026} which forecasts the constraints on $H_0$ that can be achieved from the synergy between third-generation GW networks and the SKAO. Meanwhile, this paper provides a realistic evaluation of the LVK network capabilities with both MeerKAT and SKAO, accounting for observational and instrumental limitations. See \cite{SKAObook} for a recent review of the various synergies between GW detections and SKAO.

This paper is organised as follows. Section~\ref{sec:data_generation} describes the generation of mock data consistent with our forecasted data products. Section~\ref{sec:LoS} outlines the construction of a redshift prior based on the \hi\ distribution from intensity maps. Section~\ref{sec:hierarchical_bayesian_framework} introduces the hierarchical Bayesian framework used for the analysis. We present our results in Section~\ref{sec:results} and conclude in Section~\ref{sec:conclusion},  discussing their implications along with possible extensions of this work.

\section{Data generation}
\label{sec:data_generation}

\subsection{\hi\ intensity maps}
\label{sec:data_generation_HIIM}

\begin{figure*}
    \centering

    \begin{subfigure}{0.32\linewidth}
        \centering
        \includegraphics[width=\linewidth]{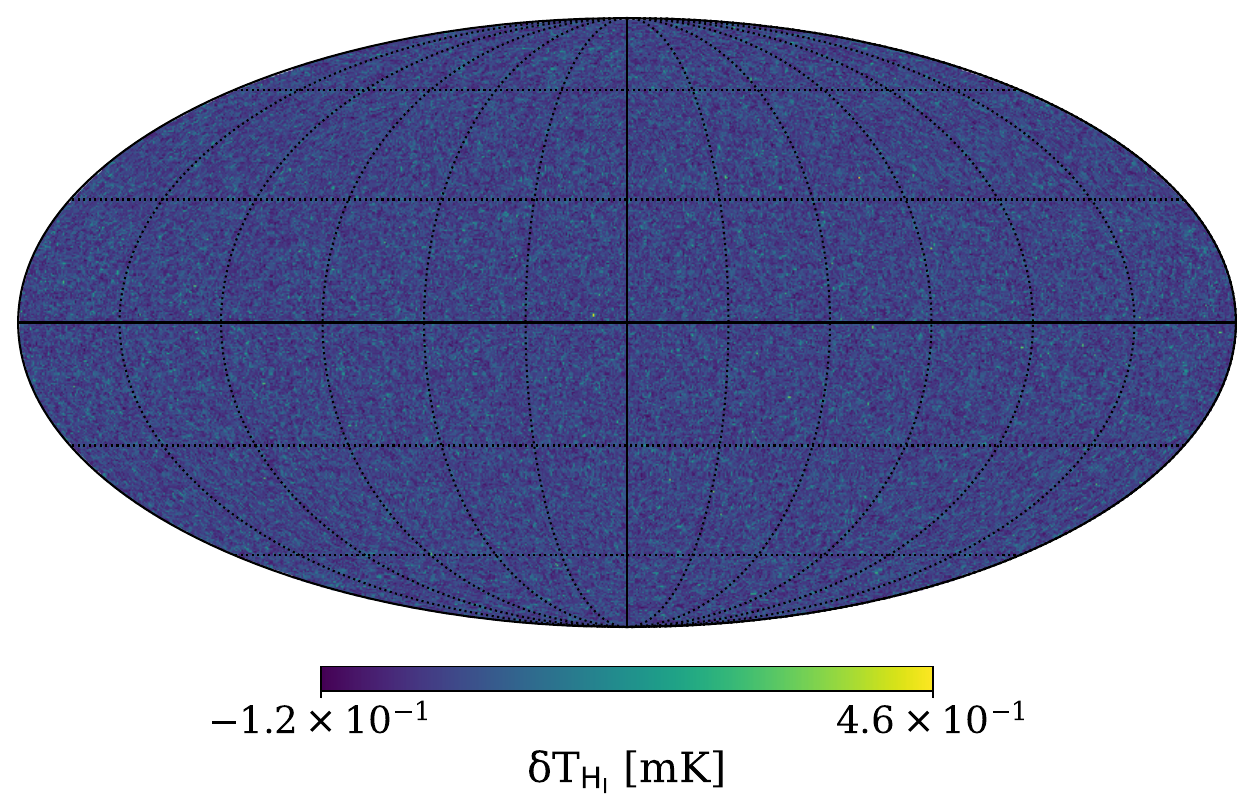}
        \caption{Original}
    \end{subfigure}
    \begin{subfigure}{0.32\linewidth}
        \centering
        \includegraphics[width=\linewidth]{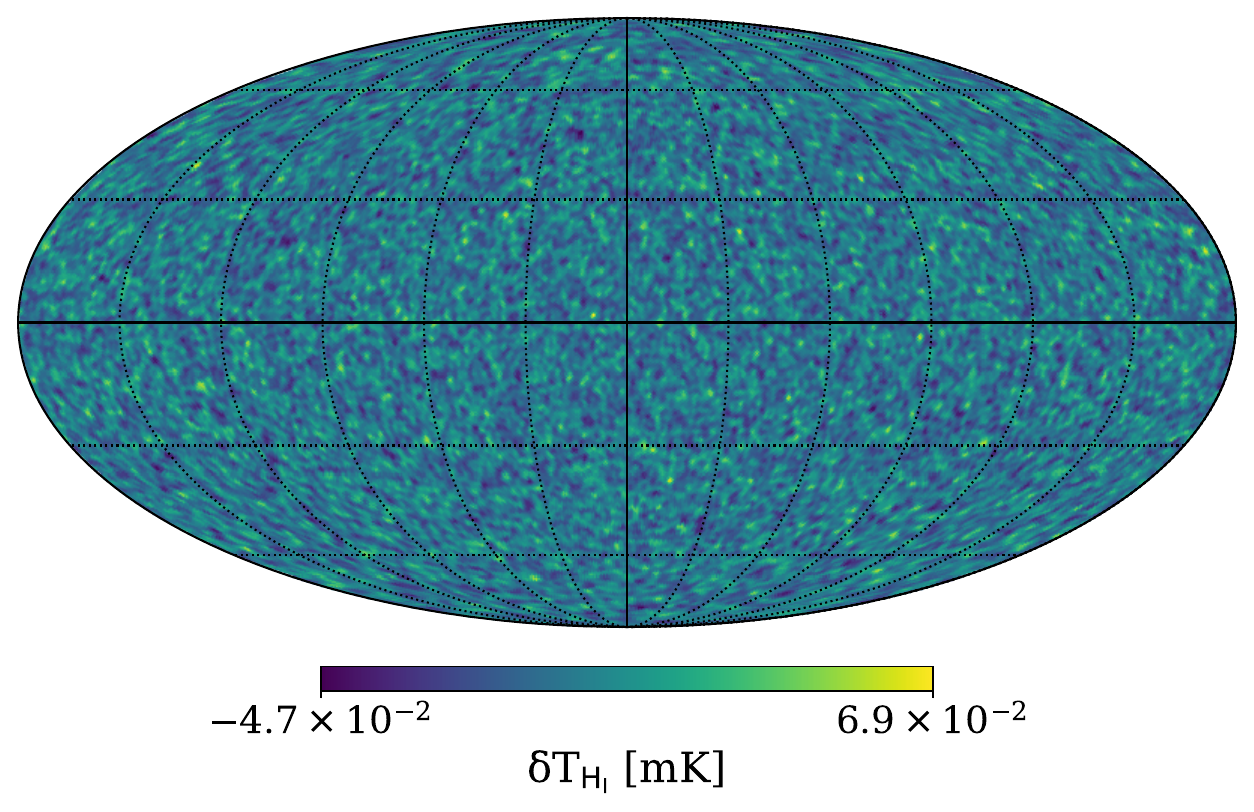}
        \caption{Beam convolution}
    \end{subfigure}
    \begin{subfigure}{0.32\linewidth}
        \centering
        \includegraphics[width=\linewidth]{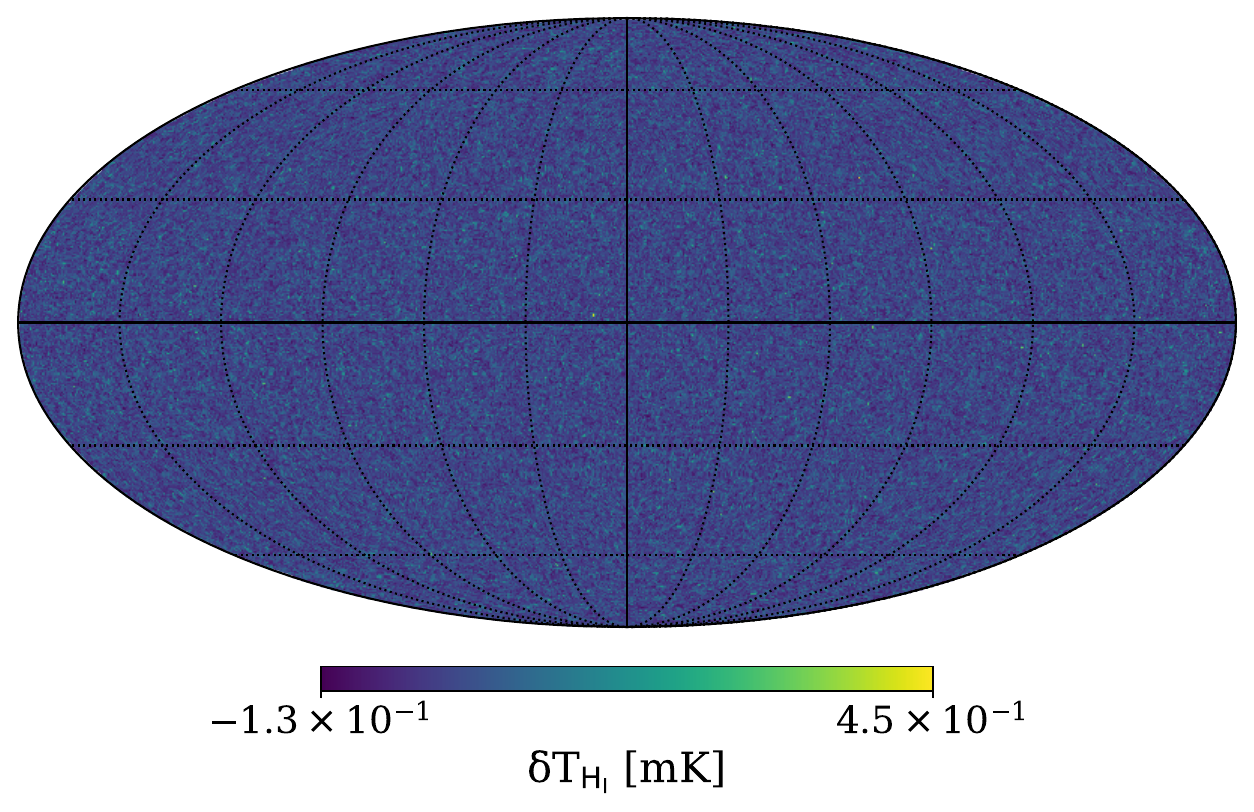}
        \caption{Instrumental noise}
    \end{subfigure}

    \medskip

    \begin{subfigure}{0.32\linewidth}
        \centering
        \includegraphics[width=\linewidth]{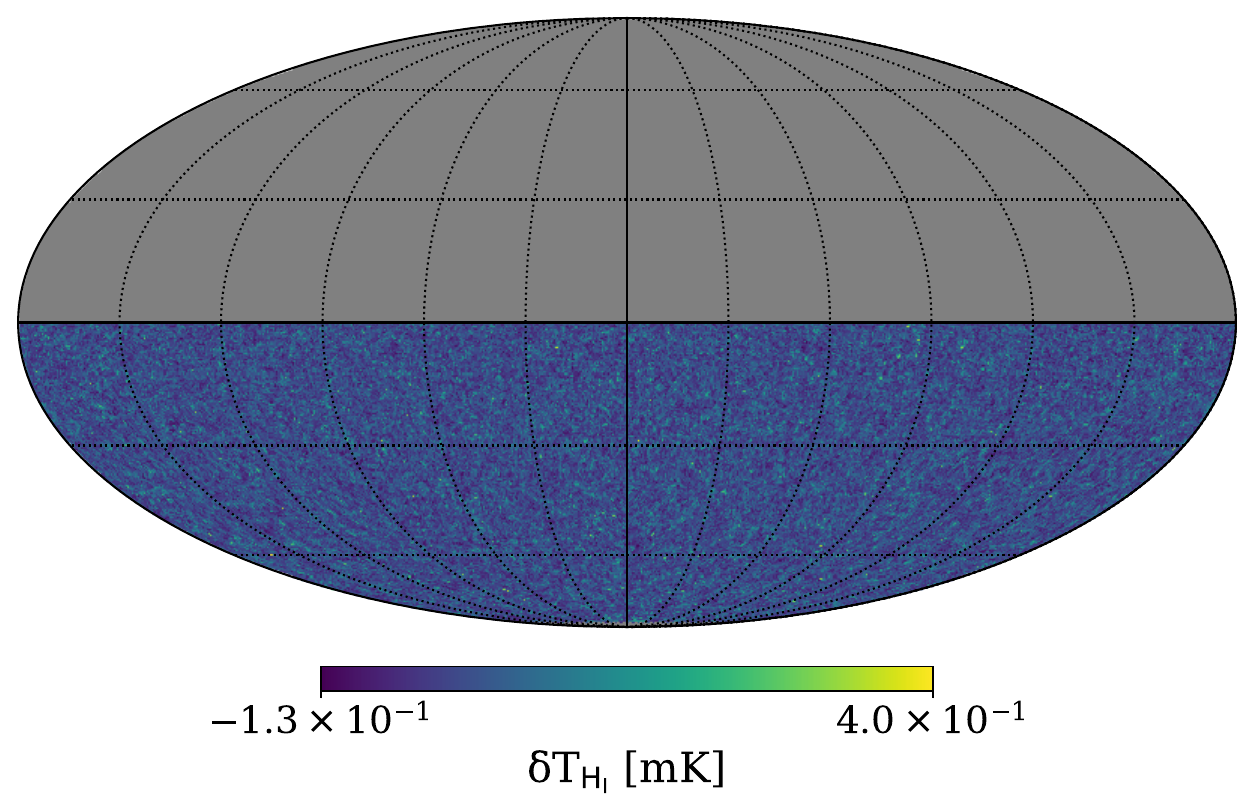}
        \caption{Partial sky coverage}
    \end{subfigure}
    \begin{subfigure}{0.32\linewidth}
        \centering
        \includegraphics[width=\linewidth]{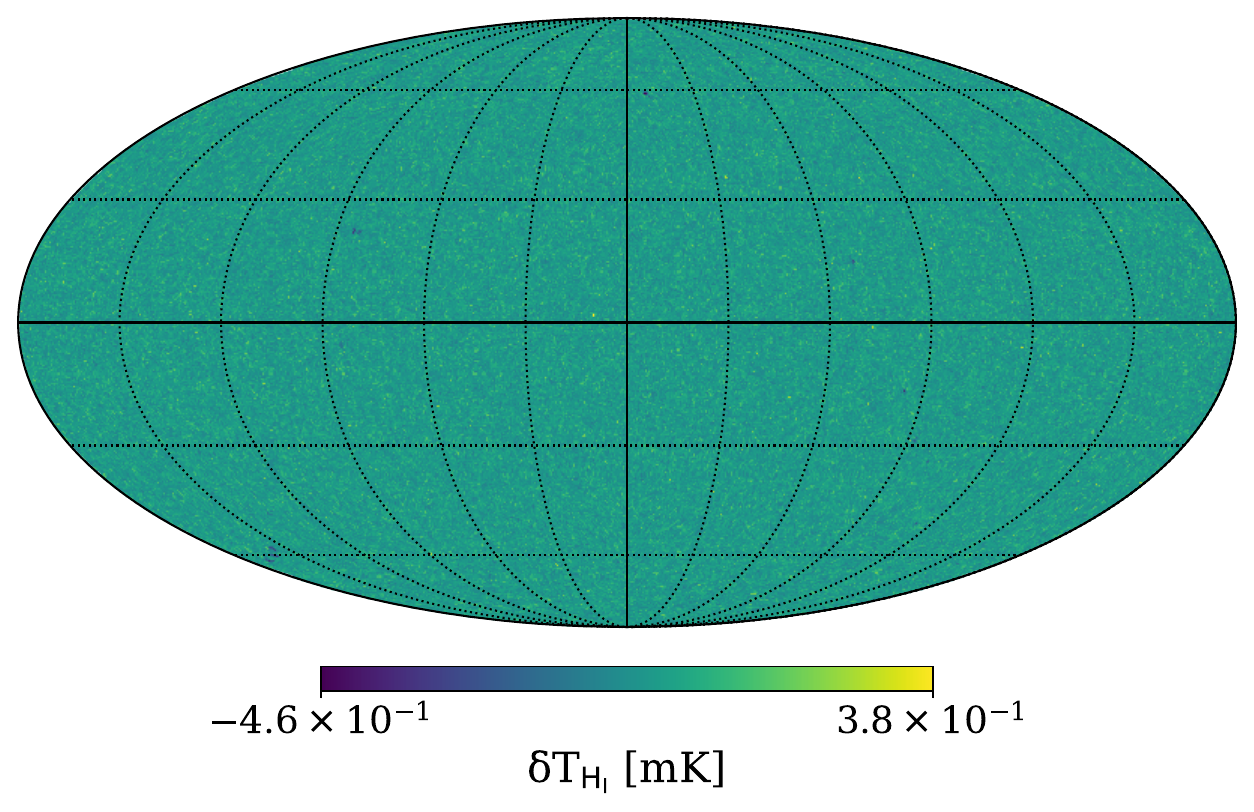}
        \caption{Foreground cleaning}
    \end{subfigure}
    \begin{subfigure}{0.32\linewidth}
        \centering
        \includegraphics[width=\linewidth]{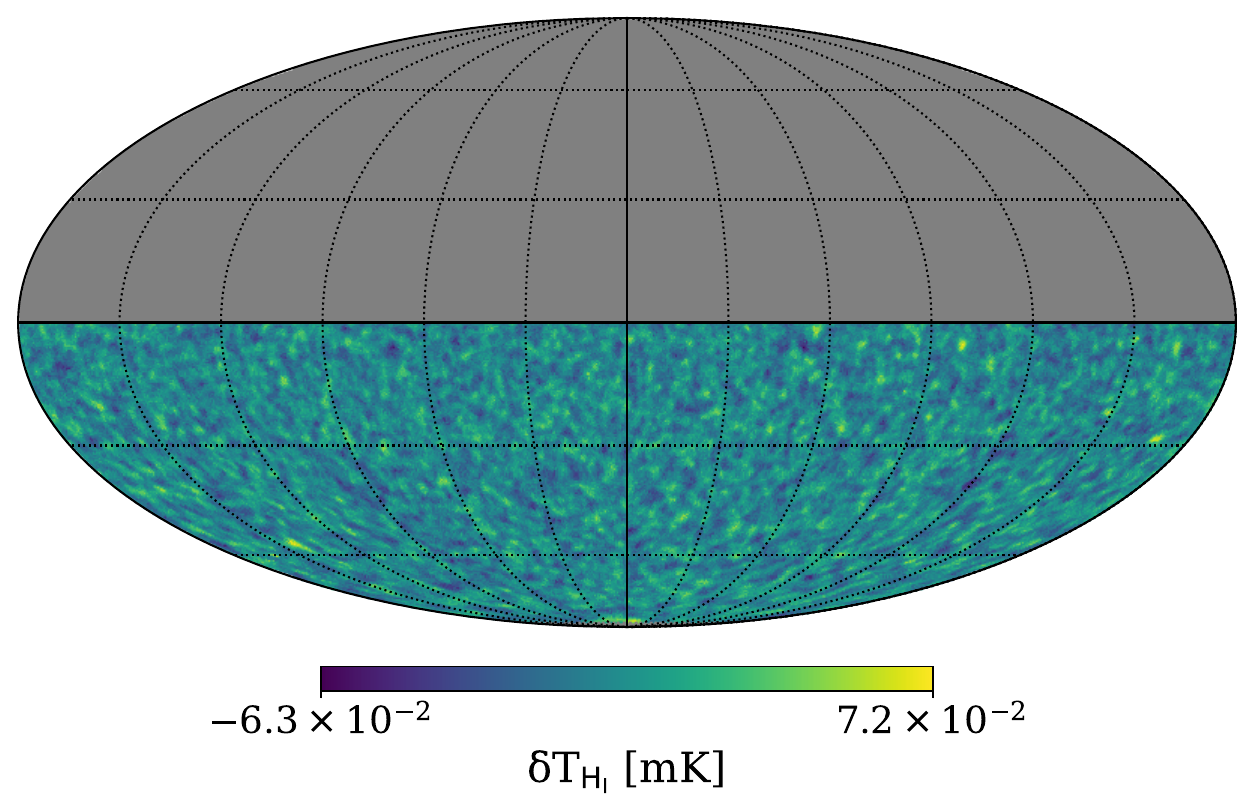}
        \caption{All effects}
    \end{subfigure}

    \caption{SKA-Mid1-like 21-cm intensity maps showing temperature fluctuations relative to the mean temperature of the field, $\delta T_{\rm H_I}$, averaged over 20 frequency bins around the effective redshift $z_{\rm eff}\approx 1.70$. We compare the ideal \hi\ map in subfigure~(a) with maps that include instrumental and observational effects: beam convolution, instrumental noise, partial sky coverage, and foreground cleaning. The impact of these effects is shown independently in subfigures~(b), (c), (d), and (e), and in combination in subfigure~(f), which therefore represents the most realistic case. Details about the survey specifications can be found in Appendix~\ref{app:parameters}.}
    \label{fig:HI_simulations}
\end{figure*}

We want to generate three-dimensional 21-cm intensity maps that trace the spatial distribution of neutral hydrogen across the sky and as a function of redshift. These maps are subsequently used to simulate the observational effects and construct realistic mock data sets, which are then jointly analysed.

We generate two independent mock intensity maps, with redshift ranges and resolutions corresponding to those of the MeerKLASS UHF-band~\citep{Santos_2017} and SKA-Mid Band 1~\citep{SKA_Cosmology_SWG_2020} surveys in single-dish mode~\citep{Battye_2013}. For simplicity, we refer to the surveys as MeerKLASS-UHF and SKA-Mid1, respectively, throughout the rest of the paper. The survey specifications are summarised in Table~\ref{tab:survey_specs}. The frequency bandwidth and channel width are particularly important parameters, as they define the redshift range and the redshift error of each layer of the map. The other parameters matter only when modelling realistic observational scenarios.
We focus on the wide MeerKLASS-UHF and SKA-Mid1 surveys because, compared with intermediate- and deep-survey configurations such as SKA-Low, SKA-Mid2, and MeerKLASS-L, they cover a larger fraction of the sky and better span the redshift range in which most LVK BBH detections occur, i.e. $z<3$ at Observing Run~4 (O4) and 5 (O5) sensitivity. This is an important advantage of \hi\ IM because, instead, stage~IV galaxy surveys remain highly incomplete at $z>1$. Most importantly, the completeness of a galaxy catalogue must be accurately characterised to avoid introducing systematic biases in dark-siren analyses, and modelling catalogue completeness as a function of redshift and galaxy properties is challenging and introduces a non-trivial dependence on additional nuisance parameters.

The 3D \hi\ power spectrum, $P_{\rm H_I}(k,z)$, describes the clustering of \hi\ by characterising the amplitude of density fluctuations as a function of scale $k$ and redshift $z$. Since \hi\ is a biased tracer of Cold Dark Matter (CDM), we write
\begin{equation}
    P_{\rm H_I}(k,z) \propto b_{\rm H_I}^2(k,z)\,P_{\rm CDM}(k,z),
\end{equation} 
where $b_{\rm H_I}(k,z)$ is the \hi-to-CDM scale- and redshift-dependent bias factor and $P_{\rm CDM}(k,z)$ is the CDM 3D power spectrum.
On sufficiently large, linear scales, the bias is constant at first approximation. Given the redshift ranges and map resolution considered in this work, this is a reasonable assumption; we therefore set $b_{\rm H_I}=1.49$ following the findings in ~\cite{Villaescusa-Navarro_2018}. With the public code \texttt{gridimp}\footnote{\url{github.com/stevecunnington/gridimp}}~\citep{gridimp}, we create \hi\ maps by drawing lognormal realisations from the \hi\ power spectrum. The resulting maps are in units of brightness temperature (mK) and are pixelated using \texttt{HEALPix}~\citep{healpix}. By default, we choose \texttt{nside=128} which corresponds to an angular resolution $\delta\theta=0.4581~\rm{deg}$, well below the current best-localised GW detection. An example is shown in Figure~\ref{fig:HI_simulations}(a). We choose \texttt{nside=128} because it is the nominal map resolution for MeerKLASS-UHF and SKA-Mid1. We further discuss the impact of this assumption on our study in Appendix~\ref{app:nside}.

However, these \textit{original} maps are not representative of the data we expect to obtain from \hi\ IM experiments. Four factors expected to limit the quality and informativeness of the mock maps considered in our analysis are introduced below. Their effect is illustrated in Figure~\ref{fig:HI_simulations}.

\begin{enumerate}
    \item \textbf{Beam convolution.} The telescope dishes are not point-like apertures but rather complex beam patterns that induce a decrease in angular resolution when the intrinsic map is convolved with the beam pattern. Following~\cite{MeerKLASS_Lband_2025}, we approximate the beam pattern to a Gaussian aperture and apply Gaussian smoothing to the map such that $\mathcal{N}(\mu=0,~\sigma=\sigma_{\rm beam})$. The Gaussian beam profile is frequency-dependent and is given by
    \begin{equation}
        \sigma_{\rm beam}(\nu)= \frac{1}{2\sqrt{2\,\rm ln(2)}} \frac{c}{\nu \, \rm D_{dish}},
        \label{eq:Gaussian_beam}
    \end{equation}
    where $\nu$ is the frequency and $\rm D_{dish}$ the diameter of the telescope dishes in meters.
    Provided that $\sigma_{\rm beam}(\nu)\lesssim \delta\theta$, where $\delta\theta$ is the angular size of the pixels as specified by \texttt{nside}, the impact on the results should be negligible. Instead, in the limit that $\sigma_{\rm beam}(\nu)\gg\delta\theta$, applying the beam convolution can result in an important decrease in structure contrast. Both SKA-Mid1 observations with $\sigma_{\rm beam}\in[0.458, 1.374]~\rm{deg}$ and MeerKLASS-UHF observations with $\sigma_{\rm beam}\in [0.515,0.916]~\rm{deg}$ are affected by this effect, which results in an important reduction in the range of temperature fluctuations especially at high redshifts (see Figure~\ref{fig:HI_simulations}). The technical specifications and observational parameters of the telescopes used to obtain these results are detailed in Appendix~\ref{app:parameters}, while Appendix~\ref{app:nside} discusses the impact of beam convolution with varying map resolution.

    \item \textbf{Instrumental noise.} The map \hi-to-noise ratio is an important consideration in 21-cm IM because the signal is extremely faint. Instrumental noise is well approximated by Gaussian noise centred on mean $\mu = 0$ and frequency-dependent standard deviation $\sigma_{\rm noise}$, derived from the radiometer equation~\citep{Dicke_1946,Campbell_2002}:
    \begin{equation}
    \label{eq:radiometer}
        \sigma_{\rm noise}(\nu)=\frac{T_{\rm sys}(\nu)}{\sqrt{2 \,\delta \nu \,t_p}},
    \end{equation}
    where $T_{\rm sys}$ sums up contributions from the receiver temperature, the spill-over noise, the CMB and the Milky-Way emissions as defined in \cite{SKA_Cosmology_SWG_2020}. For simplicity, we assume that MeerKAT and the SKAO have the same receiver temperature and adopt the expression given in Equation~10 of \cite{Cunnington_2022}. See Appendix~\ref{app:parameters} for more details on $\delta \nu$ and $t_p$. The resulting noise level is $\sigma_{\rm noise}\in[0.138, 0.199]~\rm{mK}$ for MeerKLASS-UHF and $\sigma_{\rm noise}\in[0.036, 0.121]~\rm{mK}$ for SKA-Mid1, resulting in a \hi-to-noise ratio of approximately $1-3$ and $5-8$, respectively.

    \item \textbf{Foreground cleaning.} Foregrounds such as synchrotron emission, bremsstrahlung, and bright point-like radio sources are several orders of magnitude brighter than the 21-cm signal; therefore, foreground cleaning is a major challenge in IM. Insufficient removal can leave foreground residuals, while excessive cleaning leads to signal loss, both extremes resulting in a decrease in the \hi-to-noise or \hi-to-foreground ratios, respectively. See \cite{Cunnington_2021} and \cite{Spinelli_2022} for recent reviews on foreground cleaning techniques and Appendix~\ref{app:foreground} for further details on the implementation in this work.
    
    \item \textbf{Partial sky coverage.} Finite observing times and accessible sky areas limit the fraction of the sky that can be observed, $f_{\rm sky}$. In particular, SKAO and MeerKAT observe the Southern hemisphere. To account for the incomplete survey footprints, which are an important consideration in GW cosmology, we crop the maps according to the specifications of the \hi\ IM surveys considered. See Appendix~\ref{app:parameters} for additional details on the sky fraction $f_{\rm sky}$.
\end{enumerate}

\subsection{Gravitational-Wave mock catalogue}
\label{sec:data_generation_GW}
We define a mock GW catalogue containing, for each detection, the primary and secondary source-frame masses $m_{1,\rm s}$ and $m_{2,\rm s}$, the redshift $z$, and the sky position $\phi$.  We define probability distributions to draw these quantities from, consistently with the \hi\ map, cosmology, and CBC population. We repeat the process for the MeerKLASS-UHF-like and SKA-Mid1-like maps separately. Table~\ref{tab:hyper-parameters} summarises the hyper-parameters used in the analysis.

We assume that the intrinsic CBC merger-rate density follows the same parametrised shape as the SFR function, $\Psi(z)$. The corresponding CBC population cosmological distribution per unit redshift and solid angle is therefore expressed as 
\begin{equation}
    \mathbb{P}_{\rm cosmo}(z) = \Psi(z)\,\frac{\mathrm{d}V_c(z)}{\mathrm{d}z\,\mathrm{d}\Omega},
\label{eq:Pd(z)}
\end{equation}
where $\frac{\mathrm{d}V_c(z)}{\mathrm{d}z\,\mathrm{d}\Omega}$ is the differential co-moving volume element .A time dilation term $\frac{1}{1+z}$ must be added when working in the detector frame.
We use the Madau-Dickinson model \citep{Madau-Dickinson_2014} for the SFR function:
\begin{equation}
    \Psi(z) = \Psi(z|\gamma, \kappa, z_p) =\frac{(1+z)^\gamma}{1 + \left(\frac{1+z}{1+z_p}\right)^{\gamma + \kappa}},
    \label{eq:MD}
\end{equation}
where $\gamma$ and $\kappa$ are the power law indices of the distribution below and above the peak $z_p$, respectively. The values used for $\gamma$, $\kappa$, and $z_p$ are informed by the latest LVK results~\citep{GWTC5_cosmo} and reported in Table~\ref{tab:hyper-parameters}.

The sky position is determined under the assumption that CBC mergers are more likely to originate from dense \hi\ regions. The \hi\ density is traced by the normalised temperature fluctuation $\delta _{\rm H_I} = \delta T_{\rm H_I}/\overline{T}_{\rm H_I}$, which is, in turn, proportional to the \hi\ density contrast under the assumption adopted here. $\delta T_{\rm H_I}$ is the temperature fluctuation, as shown in Figure~\ref{fig:HI_simulations}, while $\overline{T}_{\rm H_I}$ is the mean field temperature. In our simulations, $\overline{T}_{\rm H_I}$ is constant and approximately equal to 0.15~mK for MeerKLASS-UHF and 0.19~mK for SKA-Mid1. In reality, $\overline{T}_{\rm H_I}$ varies with redshift, as derived in \cite{Battye_2013}. In over-dense regions with respect to the mean density, we have $\delta T_{\rm H_I}>0$, while in under-dense regions, $\delta T_{\rm H_I}<0$.
We, therefore, model the merger probability at sky position $\phi$, given redshift bin $z$, using a power-law weighting of the \hi\ density field:
\begin{equation}
\mathbb{P}_{\rm HI}(\phi|z)\propto(1+\delta _{\rm H_I})^\epsilon
\label{eq:P(phi|z)}
\end{equation} 
where $\epsilon$ is the temperature weighting exponent describing the likelihood of CBC mergers for a given temperature contrast. Essentially, it encapsulates how the GW mergers and \hi\ IM distributions correlate at a given redshift. For simplicity, we assume this parameter to be constant and set $\epsilon =1$. We refer to $(1+\delta _{\rm H_I})^\epsilon$ as the temperature weight. Conditioned on the redshift, $P(\phi|z)$ is independent of the SFR and is derived from the \hi\ map for all pixels, hence building a CBC merger probability map. We must have $\int \mathbb{P}(\phi|z)d\phi = 1$ to normalise the sky-position probability distribution at each redshift. We draw CBC mergers from the simulated \hi\ distribution within the available redshift ranges and, as a simplifying assumption, distribute them uniformly beyond those ranges, where no \hi\ maps are available. In reality, GW sources would continue to trace the underlying matter distribution.

Finally, the probability of a CBC merger at sky position $\phi$ and redshift $z$, $\mathbb{P}(\phi,z)$, is given by
\begin{equation}
    \mathbb{P}(\phi, z) \propto \mathbb{P}_{\rm HI}(\phi|z)~\mathbb{P}_{\rm cosmo}(z),
\label{eq:P(phi,z)}
\end{equation}
and is normalised such that its sum over all voxels is unity.

\begin{figure}
    \centering
    \includegraphics[width=\linewidth]{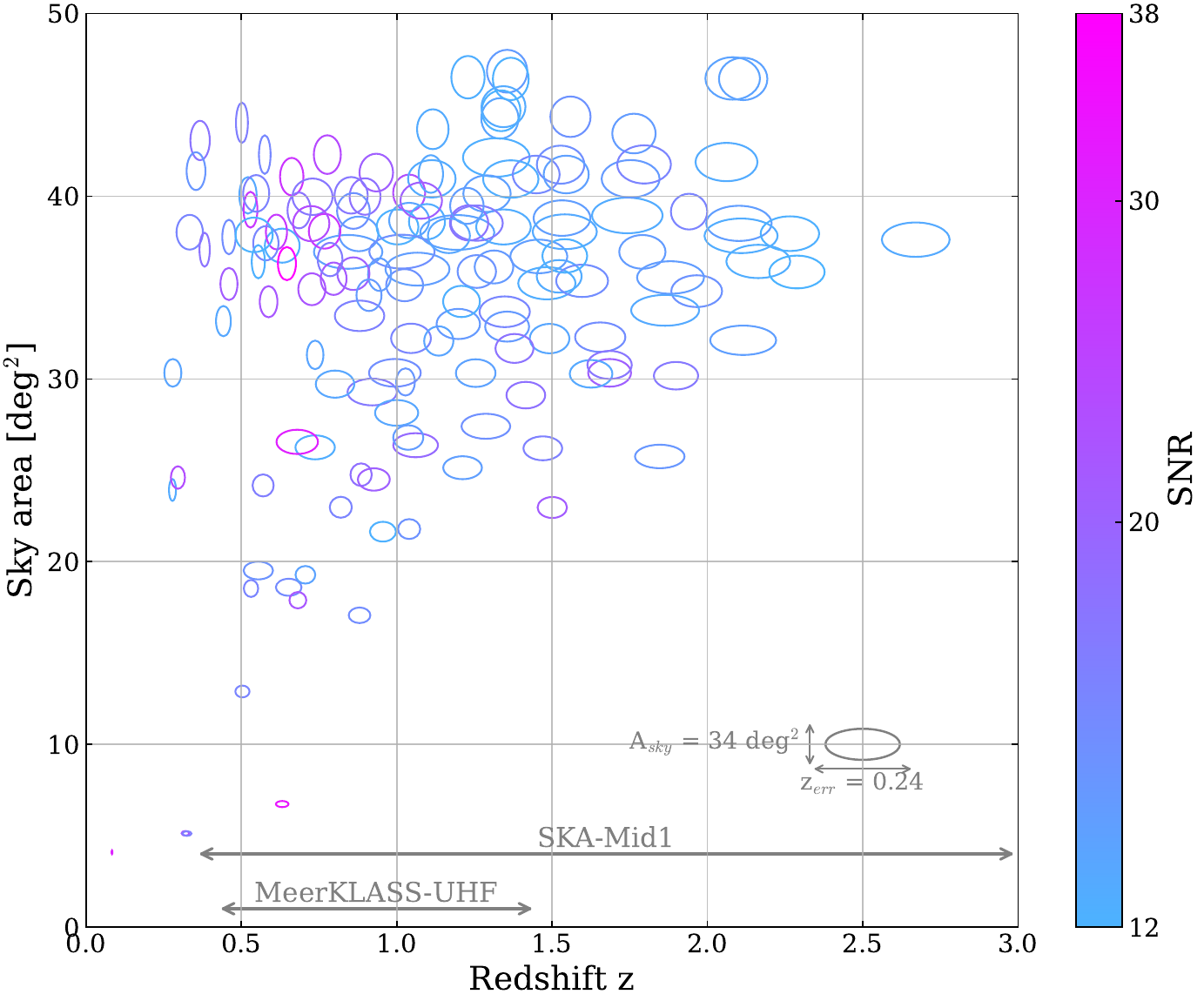}
    \caption{Distribution in redshift and sky area of the 146 SNR>12 CBC mergers drawn from the MeerKLASS-UHF-like map. Each ellipse represents a single event, with its colour encoding the event's SNR, its horizontal width proportional to the redshift error (converted from luminosity-distance error assuming Planck18 cosmology~\citep{Planck2020}), and its vertical height proportional to the 90\% credible sky area. The legend ellipse in gray corresponds to a 0.24 error in redshift and $ 34~\rm deg^2$ sky area, the mean values for our datasets, and illustrates the mapping used.}
    \label{fig:GW_z_distribution}
\end{figure}

The primary source-frame mass distribution, $\mathbb{P}_{\rm pop}(m_{\rm 1,s}|\Lambda)$, follows a power law with two Gaussian peaks, referred to as \texttt{MultiPeak} model in \cite{GWTC5_cosmo}. It is parameterised by the lower and upper limits $m_{\rm min}$ and $m_{\rm max}$ respectively, the power law index $\alpha$, the Gaussian mean values $\mu_{\rm g,low}$ and $\mu_{\rm g,high}$, the Gaussian width values $\sigma_{\rm g,low}$ and $\sigma_{\rm g,high}$, the Gaussian-peak weights $\lambda_{\rm g}$ and  $\lambda_{\rm g, low}$, and the lower-mass cut-off smoothing parameter $\delta_{\rm m}$. Instead, the secondary mass follows a simple power law with index $\beta$ between $m_{min}$ and $m_{max}$. We impose the condition $m_{\rm 1,s}>m_{\rm 2,s}$. The adopted values for the mass population parameters are informed by the LVK GWTC-5 results~\citep{GWTC5_cosmo}, as reported in Table~\ref{tab:hyper-parameters}.

The GW population is transformed into detector-frame quantities by computing the luminosity distance $d_l$ and the detector-frame masses $m_{1,\rm d}$ and $m_{2,\rm d}$, while the sky coordinates remain unchanged. The accuracy and precision with which detector-frame quantities are measured varies substantially depending on the chosen network of detectors and signal-to-noise ratio (SNR) threshold. We use \texttt{bilby}~\citep{bilby} to perform detector-frame parameter estimation. We adopt the LVK design amplitude spectral densities (ASDs) to forecast for the LVK network O5 sensitivity.\footnote{The LVK ASDs are publicly available at \url{https://dcc.ligo.org/LIGO-T2000012/public}. We adopt the A+ design target for the two LIGO detectors, the high-noise Virgo target sensitivity for O5, and the 25~Mpc KAGRA sensitivity used for high-sensitivity O4 simulations.} We set the duty cycle to 70\% for LIGO and Virgo and 60\% for KAGRA. GW signals are generated with the IMRPhenomXAS \citep{Pratten_2020}  waveform template for non-precessing BBHs. We simulate 52,000 events and use the detector network SNR as the selection criterion. Setting $\rm SNR>8$ as the detection threshold, approximately 700 GW events are detected for one year of observations at O5 sensitivity. This is consistent with the projected number of annual detections, averaged over the different phases of O5.\footnote{See \url{https://rtd.igwn.org/projects/userguide/en/latest/capabilities.html} for the projected observational capabilities of the LVK network.} For the analysis, we impose the more stringent threshold of $\rm SNR>12$, which results in a decreased number of events: 146 for MeerKLASS-UHF and 147 for SKA-Mid1. For the MeerKLASS-UHF-like (SKA-Mid1-like) catalogue, increasing the detection threshold from $\rm SNR>8$ to $\rm SNR>12$ reduces the mean redshift error by 31\% (32\%) and the mean sky-localisation area by 13\% (13\%). The median SNR consequently increases from 9.6 (9.9) to 14.6 (14.5), while the median redshift decreases from 1.49 (1.48) to 1.13 (1.05). These statistics highlight the impact of the SNR threshold and the realisation-to-realisation variability on the event selection. Figure~\ref{fig:GW_z_distribution} shows the distribution of these 146 events in redshift and sky area, as well as their SNR and redshift errors. We observe that the SKA-Mid1 survey provides near-complete coverage of the O5 LVK sensitivity range for $\rm SNR>12$ detections. MeerKLASS-UHF covers a more limited redshift range, but nevertheless provides optimal coverage of high-SNR detections, which are generally better localised and have more precise redshift measurements. Note that, given our methodology, the MeerKLASS-UHF and SKA-Mid1 results are based on two independent realisations. Consequently, their redshift, sky-area, and SNR distributions may differ, and these differences in event selection may affect the cosmological inference. Future work will consider changing methodology to draw events from a single \hi\ field, resulting in a common GW catalogue.

\section{Constructing a \hi-informed line-of-sight redshift prior}
\label{sec:LoS}
\begin{figure}
    \centering
        \includegraphics[width=\linewidth]{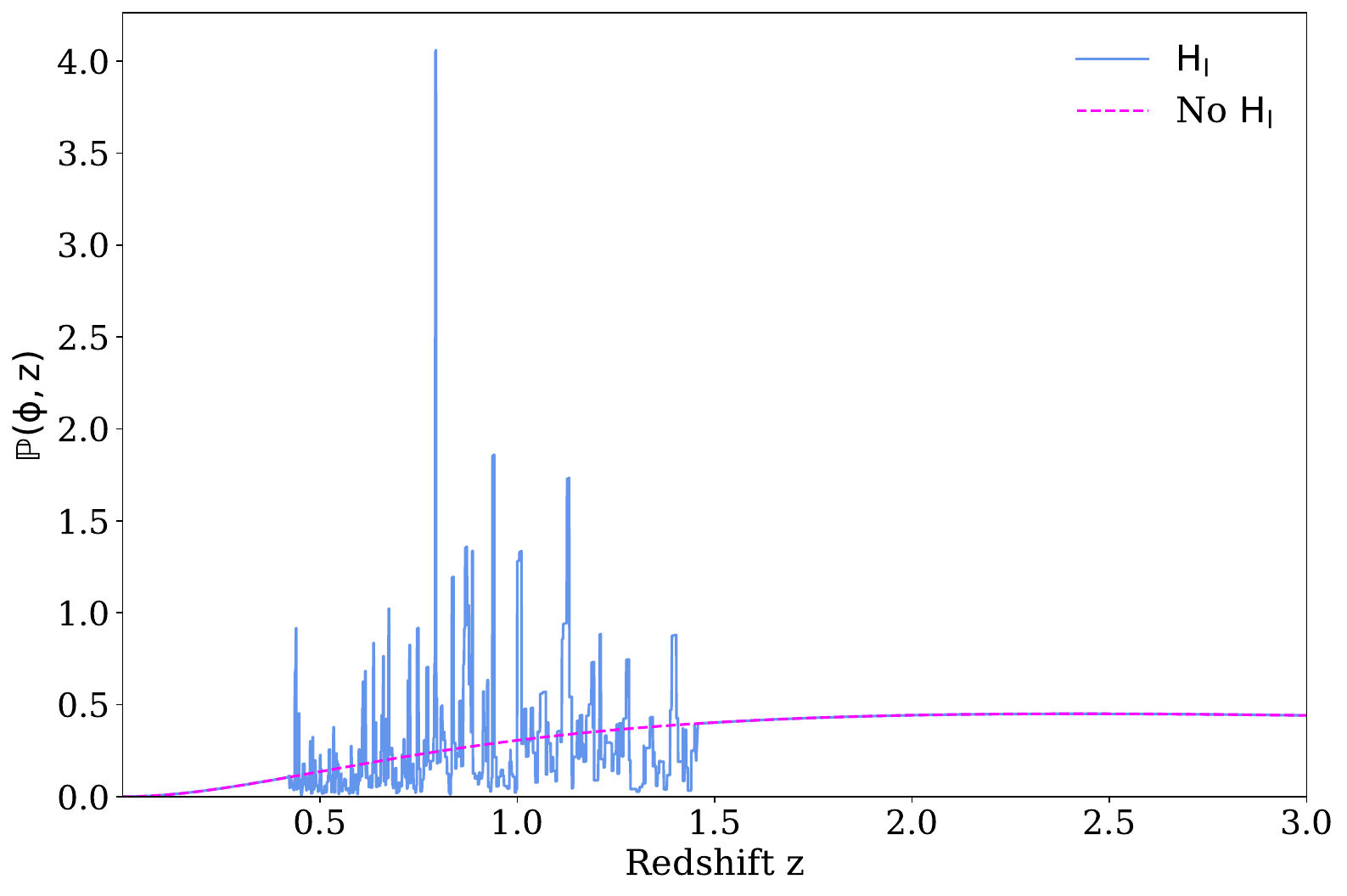}
    \caption{Line-of-sight redshift priors for a single pixel informed by the MeerKLASS-UHF \hi\ temperature map. The dashed magenta line shows the redshift prior of an homogeneous field, ie. without \hi\ while the blue line shows the redshift prior constructed from the \hi\ temperature fluctuations.}
    \label{fig:LoS}
\end{figure}

\begin{figure}
    \centering
        \includegraphics[width=\linewidth]{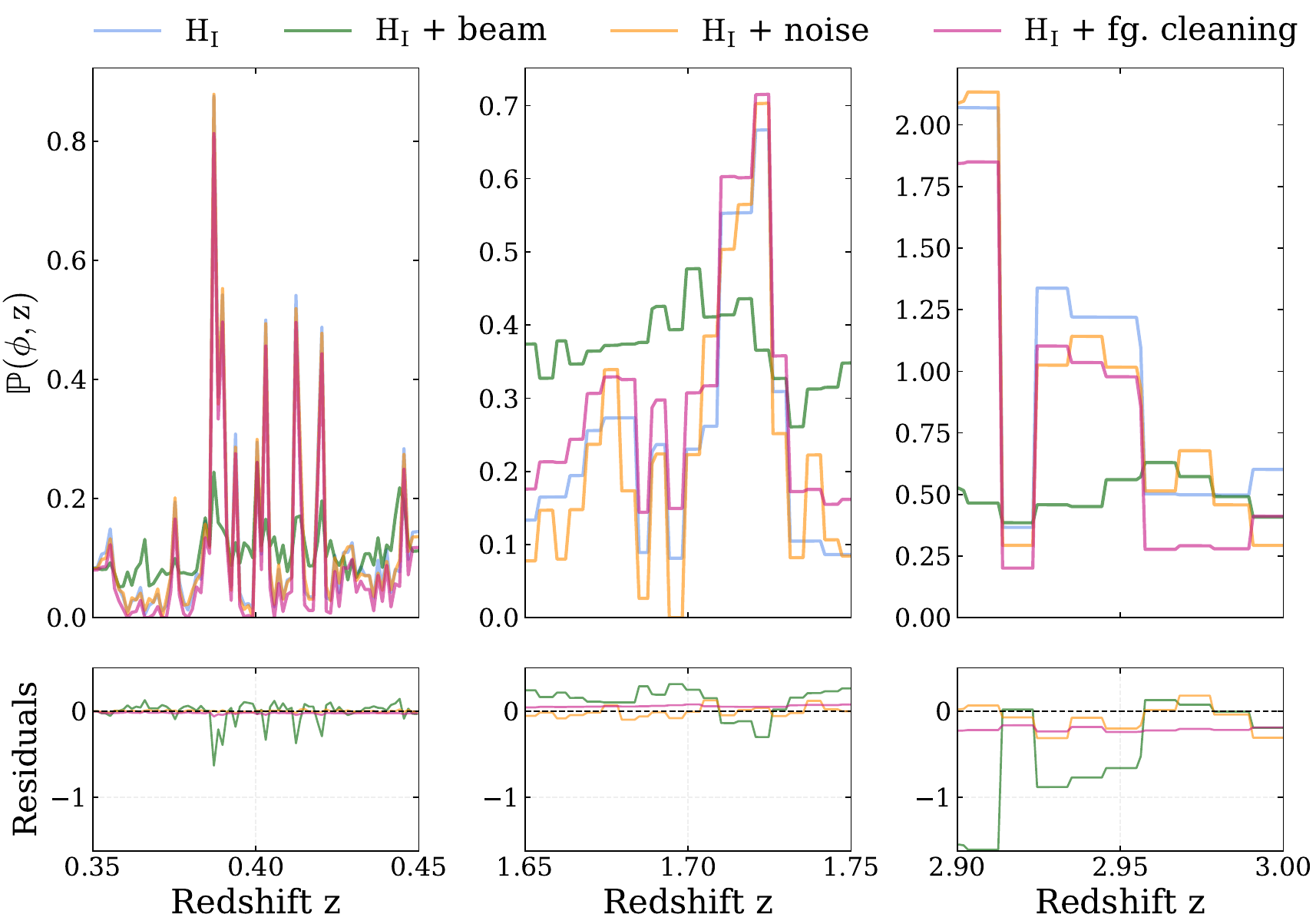}
    \caption{Impact of beam convolution, instrumental noise and foreground cleaning on the single-pixel line-of-sight redshift priors informed by the SKA-Mid1 \hi\ temperature map. The three vertical panels show distinct narrow redshift intervals, the left and right ones at the SKA-Mid1 redshift boundaries, and the middle one centred on the effective redshift (see Table~\ref{tab:survey_specs}). The top row shows the line-of-sight redshift priors, while the bottom row shows the residuals with respect to the original \hi\ map.}
    \label{fig:LoS_effects}
\end{figure}

In the dark-siren framework, the redshift prior of each GW detection is informed by the distribution of the chosen EM tracer along the line of sight. We consider the pixels $\phi_i$ within the 90\% sky area of each GW event, i.e. $i=\{1,N_{\rm skymap}\}$, and construct for each of them a line-of-sight redshift prior, $\mathbb{P}(z|\phi_i)$, by summing the CBC merger probability in all redshift bins $z_j$, i.e. $j=\{1,N_z\}$ such that
\begin{equation}
    \mathbb{P}_{\rm LoS}(z) \approx \sum_{i=1}^{N_{\rm skymap}}\mathbb{P}(z|\phi_i)  \approx \sum_{i=1}^{N_{\rm skymap}}\sum_{j=1}^{N_{z}}\mathcal{U}(z_j, z_{j+1})\frac{\mathbb{P}(\phi_i,z_j)}{\mathbb{P}(\phi_i)}.
\label{eq:LoS}
\end{equation}
$\mathbb{P}(\phi_i,z_j)$ is defined in Equation~\ref{eq:P(phi,z)}, while $\mathbb{P}(\phi_i)$ corresponds to the skymap probability distribution. A time dilation term $\frac{1}{1+z_j}$ must be added when working in the detector frame. Finally, $\mathcal{U}(z_j, z_{j+1})$ is the uniform redshift error function. In \hi\ IM, the redshift uncertainty arises from the finite frequency resolution of the surveys. The uniformly-spaced frequency bins $\delta \nu$ result in non-uniform binning in redshift such that $\delta z = |\frac{\delta \nu}{d\nu / dz}| = \frac{\delta \nu}{\nu_0} (z+1)^2$. Hence, we define the redshift error function as a uniform function, $\mathcal{U}(z_j, z_{j+1})$, defined between $z_j$ and $z_{j+1}=z_j+\sigma_j$, where $\sigma_j=\delta z(z_j)$. This is clearly illustrated in Figures \ref{fig:LoS} and \ref{fig:LoS_effects}, where the high-$z$ bins are substantially wider. We use a frequency resolution of 1~MHz, whereas the MeerKLASS-UHF and SKA-Mid1 receivers actually achieve a resolution of 0.1~MHz, corresponding to redshift bins ten times narrower. Our choice is therefore conservative, and the redshift precision quoted here should be regarded as a lower limit on what is actually achievable. 

Figure \ref{fig:LoS} compares the line-of-sight redshift prior for a single pixel from an inhomogeneous MeerKLASS-UHF-like field with that of a homogeneous field, which essentially follows the co-moving volume element $\frac{\mathrm{d}V_c(z)}{\mathrm{d}z\,\mathrm{d}\Omega}$. 
The blue line, constructed from \hi\ temperature fluctuations, shows significant structure indicative of variations in the \hi\ density along the redshift direction. The collective structure from combining all lines of sight contributing to an event's estimated sky area provides additional information on the relative probability of the source being located at different redshifts. In contrast, the magenta dashed line which is uninformative and is effectively equivalent to the spectral-siren scenario. Outside the \hi\ field redshift coverage, the redshift prior is set to that of the homogeneous field. This would also apply to discarded redshift bins in the case that they are removed due to additional observational limitations, such as radio frequency interferences, which may require the removal of contaminated frequency channels.

Figure \ref{fig:LoS_effects} shows how the observational effects considered in Section~\ref{sec:data_generation_HIIM} modify the intrinsic redshift prior at low, medium, and high redshift. 
As previously observed in Figure~\ref{fig:HI_simulations}, beam convolution reduces the contrast of the redshift prior relative to the ideal case by suppressing small-scale structure. It can also introduce spurious features in the line-of-sight prior of a single pixel, because emission from bright neighbouring pixels leaks into fainter ones. This is of limited concern for dark-sirens analysis: combining the pixels within an event’s localisation region averages over neighbouring lines of sight. Instrumental noise adds random fluctuations to the redshift prior that are not associated with the underlying \hi\ distribution. The noise level in SKA-Mid1 is low enough that these stochastic fluctuations do not exceed the \hi\ structure and are not mistakenly interpreted as a dominant physical signal. As discussed in Section~\ref{sec:data_generation_HIIM}, the noise level is much higher in the MeerKLASS-UHF scenario, leading to a less robust redshift prior when noise is included, relative to the original map. The impact of instrumental noise on the redshift prior therefore depends on its amplitude relative to the \hi\ signal. Foreground cleaning modifies the redshift prior in a more complex manner while mostly preserving the features of the original \hi\ map. The foreground removal method used in this study removes the dominant, spectrally smooth modes in the frequency-frequency covariance, which are associated primarily with foreground emission. Because some \hi\ signal also overlaps with these modes, the cleaning can remove genuine signal, particularly broad spectral features, and distort the redshift prior. See Appendix~\ref{app:foreground} for additional details. Finally, as in Figure~\ref{fig:LoS}, the width of the redshift bins increases with redshift, following the corresponding increase in redshift error. In addition to the increasing redshift uncertainty, we find that the effects of observational limitations become increasingly pronounced at higher redshift as shown by the residuals in Figure~\ref{fig:LoS_effects}. However, this effect is likely overestimated in our forecasts due to the coarser frequency resolution adopted in the simulations. 

These modifications to the line-of-sight prior directly propagate into the $H_0$ inference. The redshift prior improves the constraints by assigning greater weight to redshifts where the GW source is more likely to lie, as indicated by the high-temperature \hi\ voxels. When combined with the GW event luminosity-distance estimate, this weighting provides statistical redshift information that replaces an electromagnetic counterpart or an individual host-galaxy identification. Beam smoothing and noise can reduce the prior's informative structure, thereby weakening the $H_0$ constraints. However, changes to the prior can also introduce biases: for example, removing \hi\ modes may distort its redshift structure and systematically alter the inferred redshift weighting. Thus, observational effects can both broaden the constraints by reducing the available information and shift the inferred $H_0$ by distorting the LoS prior.

\section{Hierarchical Bayesian framework}
\label{sec:hierarchical_bayesian_framework}
We use a Bayesian hierarchical framework to derive $p(\Lambda|\{x\})$, the posterior distribution on hyper-parameters $\Lambda$ given a set of data $\{x\}$. We rely on the formulation of a prior distribution $p(\Lambda)$ describing our initial knowledge of $\Lambda$ and a likelihood function $\mathcal{L}(\{x\}|\Lambda)$ that indicates the probability of measuring  $\{x\}$ given $\Lambda$. They are related by
\begin{equation}
    p(\Lambda|\{x\}) \propto \mathcal{L}(\{x\}|\Lambda) \, p(\Lambda).
\end{equation}
Essentially, inferring cosmological hyper-parameters consists in obtaining the posterior distributions $p(\Lambda|\{x\})$. Since we adopt uniform priors over the ranges given in Table~\ref{tab:hyper-parameters}, the posterior is proportional to the likelihood within these bounds, and its peak therefore correspond to the parameter values that maximise the likelihood. Following the framework derived in \cite{Mandel_2019} and \cite{Vitale_2021} and implemented in \texttt{icarogw}~\citep{icarogw}, the likelihood function is given by:
\begin{equation}
    \mathcal{L}(\{x\}|\Lambda) \propto e^{-N_{\rm exp}(\Lambda)} \,T_{\rm obs}\prod_{i=1}^{N_{\rm obs}} \int \mathcal{L}(x_i|\theta, \Lambda) \frac{\mathrm{d}N_{\rm CBC}(\Lambda)}{\mathrm{d}\theta\,\mathrm{d}z \,\mathrm{d}\Omega\, \mathrm{d}t} \mathrm{d}\theta\,\mathrm{d}z \,\mathrm{d}\Omega,
    \label{eq:likelihood}
\end{equation}
where $N_{\rm exp}(\Lambda)$ is the expected number of detections given the population parameters $\Lambda$, $N_{\rm obs}$ is the observed number of detections during the observation time $T_{\rm obs}$,  $\mathcal{L}(x_i|\theta, \Lambda)$ is the likelihood of observing the data $x_i$ for an individual event, given the source-frame parameters $\theta = \{m_1, m_2, \ldots\}$ and population parameters $\Lambda$, and $\frac{\mathrm{d}N_{\rm CBC}(\Lambda)}{\mathrm{d}\theta\,\mathrm{d}z\,\mathrm{d}\Omega \,\mathrm{d}t}$ is the differential detector-frame CBC merger rate per unit source parameter $\theta$ per redshift $z$ per solid angle $\Omega$.

In this work, the CBC merger rate is parameterised as follows:
\begin{equation}
    \frac{\mathrm{d}N_{\rm CBC}}{\mathrm{d}\theta\,\mathrm{d}z\,\mathrm{d}\Omega\,dt} = \frac{\mathrm{d}N_{\rm CBC}}{\mathrm{d}N_{\rm H_I}\,\mathrm{d}\theta\,\mathrm{d}t_s} \frac{\mathrm{d}t_s}{\mathrm{d}t} \frac{\mathrm{d}N_{\rm H_I}}{\mathrm{d}z\,\mathrm{d}\Omega},
    \label{eq:CBCrate1}
\end{equation}
where $N_{\rm H_I}$ is a discrete quantity that denotes the number of \hi\ voxels, introduced as a result of the pixelization of the continuous intensity map rather than reflecting physically countable objects. We discuss the impact of varying the resolution of the intensity maps in Appendix~\ref{app:nside}.  We convert the source-frame merger rate to the detector-frame rate using the time dilation relation $\frac{\mathrm{d}t_s}{\mathrm{d}t} = \frac{1}{1+z}$, where $t_s$ and $t$ are the source-frame and detector-frame times, respectively. The source-frame merger rate per \hi\ voxel per unit source parameter can be further expressed as
\begin{equation}
    \frac{\mathrm{d}N_{\rm CBC}}{\mathrm{d}N_{\rm H_I}\,\mathrm{d}\theta\,\mathrm{d}t_s} = R_{\rm H_I}(z)\, \mathbb{P}_{\rm pop}(\theta|\Lambda). 
\end{equation}
$\mathbb{P}_{\rm pop}(\theta|\Lambda)$ is the source-frame population model normalised such that $\int\mathbb{P}_{\rm pop}(\theta|\Lambda) \mathrm{d}\theta = 1$. The source-frame merger rate per \hi\ voxel, $R_{\rm H_I}(z)$, can be decomposed as
\begin{equation}
    R_{\rm H_I}(z)=R_{\rm H_I,0}\,\frac{\mathrm{d}V_c(z)}{\mathrm{d}N_{\rm H_I}}\,f(\delta _{\rm H_I})\,\Psi(z).
\end{equation}
$R_{\rm H_I,0}$ is the source-frame merger rate per co-moving volume at redshift $z=0$. $\frac{\mathrm{d}V_c(z)}{\mathrm{d}N_{\rm H_I}}$ is the co-moving voxel number density, such that $R_{\rm H_I,0}\,\frac{\mathrm{d}V_c(z)}{\mathrm{d}N_{\rm H_I}}$ is the source-frame merger rate per \hi\ voxel. $f(\delta _{\rm H_I})$ is the temperature weight, a function of the normalised \hi\ temperature fluctuation $\delta _{\rm H_I}$ which modulates how the CBC merger rate varies with the \hi\ temperature. We use a power law parameterisation described in  Equation~\ref{eq:P(phi|z)}, that is $f(\delta _{\rm H_I})=(1+\delta _{\rm H_I})^\epsilon$. $\Psi(z)$ is the source-frame SFR that defines how the merger rate evolves with redshift, as described by Equation~\ref{eq:MD}. It must be normalised such that $\Psi(0)=1$.  The number of \hi\ voxels per redshift per solid angle can be further decompose as a function of the co-moving volume element:
\begin{equation}
    \frac{dN_{\rm H_I}}{dz\,d\Omega} = \frac{dN_{\rm H_I}}{dV_c} \frac{dV_c}{dz\,d\Omega}.
\end{equation}
Eventually, Equation~\ref{eq:CBCrate1} can be rewritten as
\begin{equation}
    \frac{\mathrm{d}N_{\rm CBC}}{\mathrm{d}z\,\mathrm{d}\Omega\,\mathrm{d}\theta\,dt} = R_{\rm H_I,0}\,\Psi(z)\,f(\delta _{\rm H_I})\,\mathbb{P}_{\rm pop}(\theta|\Lambda)\,\frac{1}{1+z} \frac{\mathrm{d}V_c(z)}{\mathrm{d}z\,\mathrm{d}\Omega}.
    \label{eq:CBCrate2}
\end{equation}
When marginalising over the source parameters $\theta$ and time $t$, the CBC merger rate described in Equation~\ref{eq:CBCrate2} is equivalent to the expression for the probability of a GW merger at redshift $z$ and sky position $\phi$, $P(\phi,z)$, described in Equation~\ref{eq:P(phi,z)}, and subsequently used to construct redshift priors.

We evaluate the likelihood using Monte Carlo integration. For each GW event, the integral over the source parameters is approximated as a Monte Carlo sum over 2000 posterior samples from the event's parameter estimation, with each sample weighted by the ratio of the population model to the prior used in the parameter estimation, $p_{\rm pop}(\theta|\Lambda)/\pi(\theta)$. The selection term $N_{\rm exp}(\Lambda)$ is estimated from a set of injections. We refer to \cite{icarogw} for implementation details. 
We sample the posterior with the \texttt{emcee} MCMC sampler~\citep{Foreman-Mackey_2013} as implemented in \texttt{bilby}~\citep{bilby}. We assess convergence using the integrated autocorrelation time ($N_{\rm steps} > 50\,\tau_{\max}$), the Gelman--Rubin statistic ($\hat{R} < 1.01$), the effective sample size ($\mathrm{ESS} > 400$), and the stability of the posterior between the two halves of the chains. Once these criteria are met, we discard a burn-in of $3\tau_{\max}$ steps per walker and use the remaining samples to construct the posterior distributions.

\begin{figure*}
    \centering
    \includegraphics[width=\linewidth]{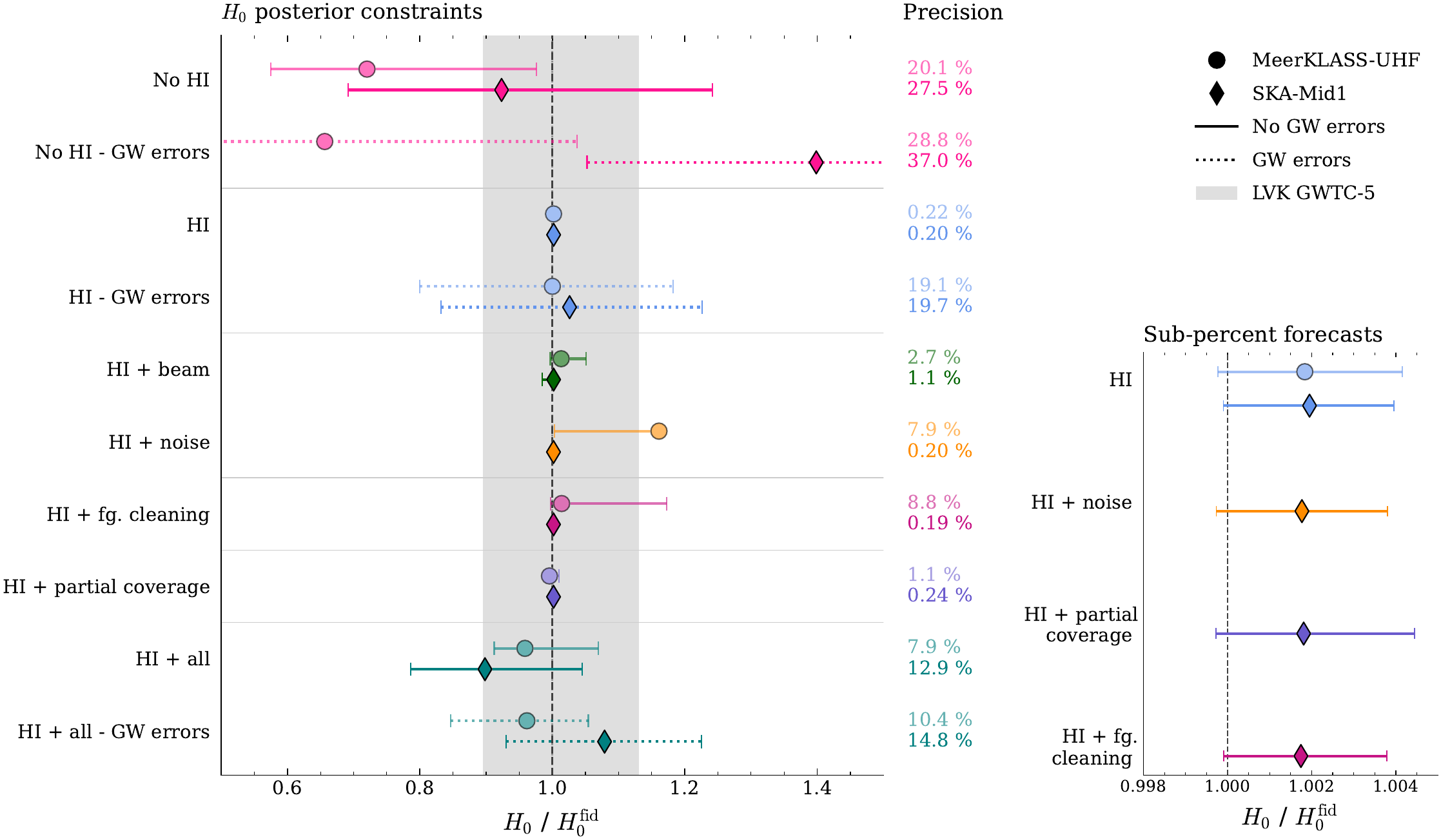}
    \caption{
    Constraints on the Hubble constant, $H_0$, obtained from hierarchical analyses combining GW observations with \hi\ information for the MeerKLASS-UHF (circular markers) and SKA-Mid1 (diamond-shaped markers) surveys. The main panel shows the posterior median and $68\%$ credible interval for each configuration and observational scenario, expressed relative to the fiducial value, $H_{0, \rm fid}$, while the percentages in the right-hand panel give the corresponding fractional precision, $\sigma_{H_0}/H_0$. We fix $\Omega_m$ for MeerKLASS-UHF's \textit{\hi\ + noise} scenario. The shaded vertical band indicates the $68\%$ credible interval associated with the LVK GWTC-5 results~\citep{GWTC5_cosmo}, relative to the median value. The right panel highlights the configurations for which sub-percent precision is obtained. The \textit{\hi} case illustrates the maximal constraining power of the method in the absence of observational limitations, while incorporating beam effects, instrumental noise, incomplete sky coverage, and foreground-cleaning allows us to construct increasingly realistic forecasts of the \hi\ information available to the hierarchical analysis. The dashed lines correspond to scenarios where the error on the GW parameter measurements are considered.
    }
    \label{fig:H0_posteriors_forest}
\end{figure*}

\begin{figure}
    \centering
    \includegraphics[width=\linewidth]{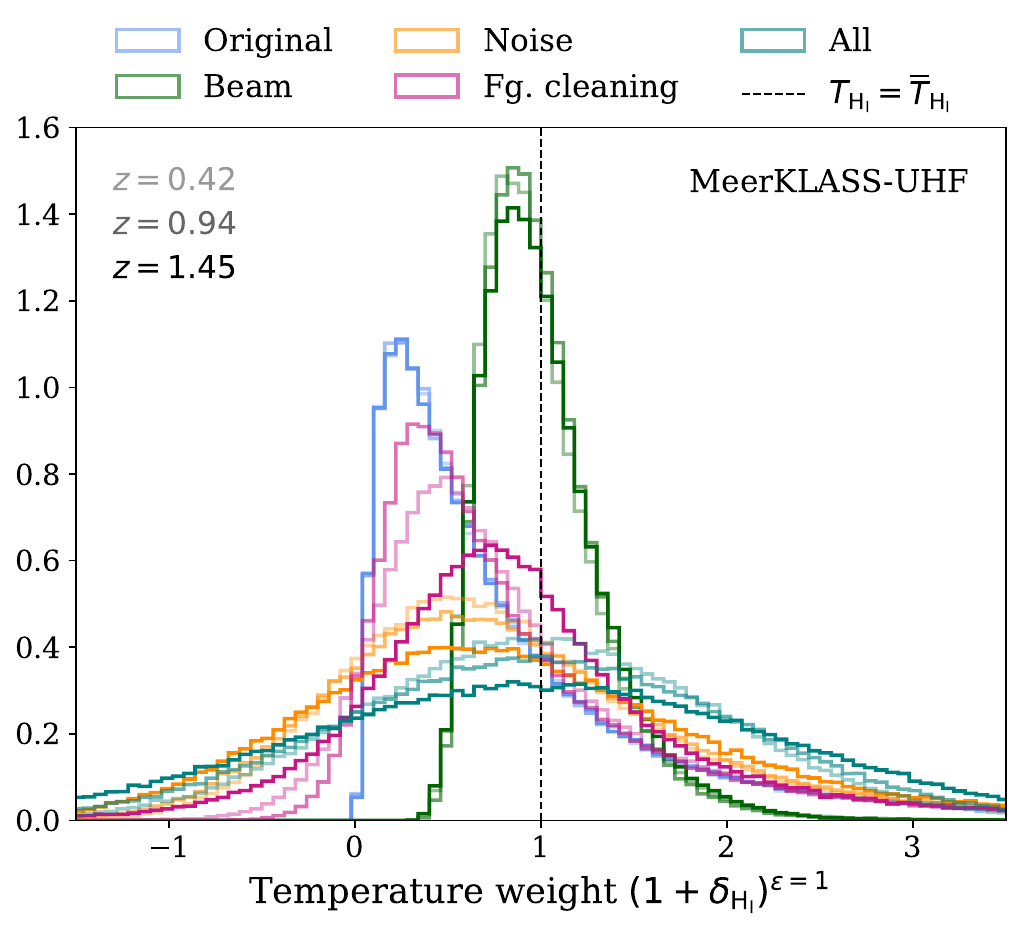}
    \includegraphics[width=1.02\linewidth]{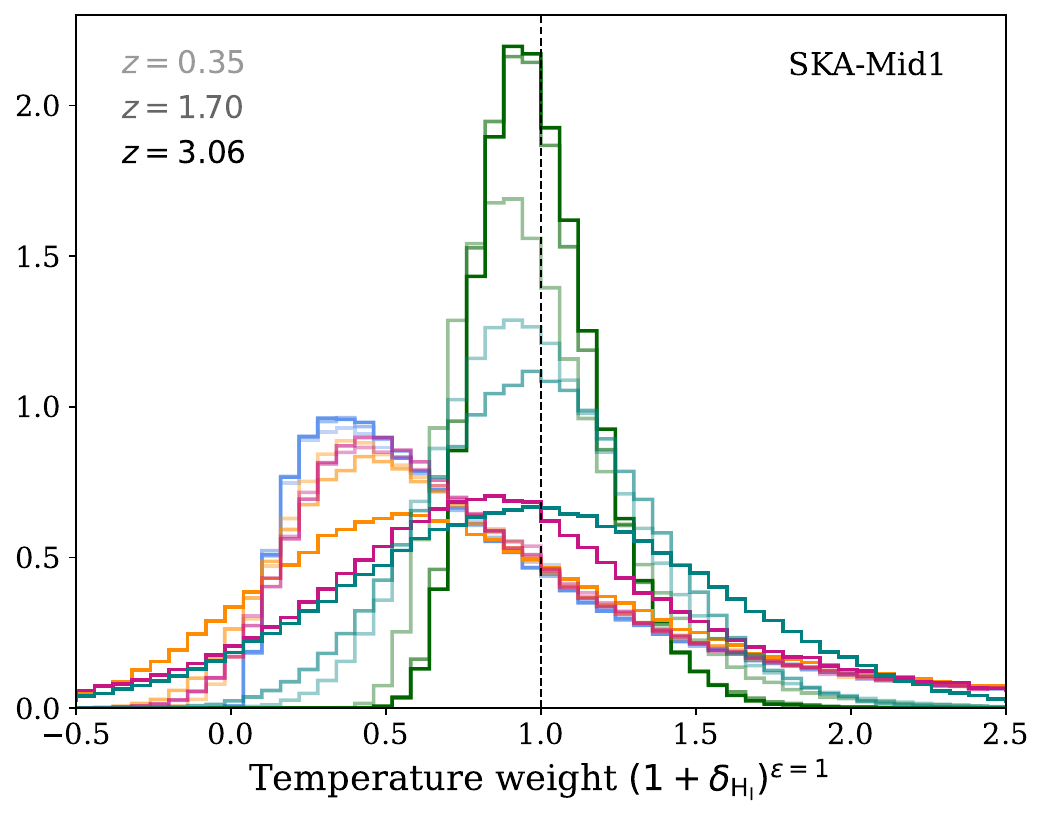}
    \caption{Temperature-weight distributions for MeerKLASS-UHF (\textit{top}) and SKA-Mid1 (\textit{bottom}). We show the distributions obtained from the original, beam-convolved, noisy, and foreground-cleaned \hi\ intensity maps, as well as from a map including all observational effects. Partial sky coverage is not included, as the distributions are normalised. Increasingly opaque lines correspond to increasingly large redshifts, with the lightest and darkest curves corresponding to the low-end and high-end redshift limits, respectively, and the intermediate curve to the effective redshift bin. The dashed line indicates a temperature weight of unity, corresponding to $\delta T_{\rm HI}=0$ and hence to a local pixel temperature equal to the mean temperature. Pixels to the left (right) of the line are under-dense (over-dense).}
    \label{fig:hist}
\end{figure}

\section{Results}
\label{sec:results}
We present results for the different observational scenarios with both MeerKLASS-UHF and SKA-Mid1, accounting for both the observational limitations affecting the \hi\ maps (see Section~\ref{sec:data_generation_HIIM}) and the uncertainties in the measured GW masses, sky positions, and luminosity distances. Figure~\ref{fig:H0_posteriors_forest} shows the median and 68\% confidence intervals resulting from Bayesian inference. Results are based on a single simulation realisation and should therefore be interpreted in light of potential realisation-to-realisation variability. Evaluating additional realisations is left for future work. Figure~\ref{fig:hist} shows the distributions of the temperature weights, $(1+\delta_{\rm H_I})^{\epsilon=1}$, across all pixels in the redshift bins $z_{\rm min}$, $z_{\rm eff}$, and $z_{\rm max}$, and how these distributions transform under the observational scenarios considered. These weights enter the construction of the line-of-sight redshift prior (see Equations~\ref{eq:P(phi|z)} and~\ref{eq:LoS}), and therefore provide a way to interpret the cosmological inference with map-level considerations.
The original \hi\ temperature-weight distribution is asymmetric, with a peak below unity and an extended tail towards high weights. The peak below unity indicates that low-weight, under-dense pixels are more common, while the high-weight tail reflects the relatively small number of over-dense pixels. This skewed shape is consistent with the lognormal realisation of the \hi\ density field used to construct the mocks and reflects the skewed nature of the true cosmological density distribution. The distribution changes only slightly across the redshift range considered, suggesting that the overall contrast of the \hi\ field is broadly similar in these bins. This follows from the mean \hi\ field temperature, $\overline{T}_{\rm H_I}$, being redshift independent.

The distribution’s peak, width, and shape help to understand how observational effects alter the maps. For example, structure suppression can narrow the distribution and reduce pixel-to-pixel contrast, making the intensity map less effective at distinguishing likely from unlikely locations along the line of sight. A broader distribution indicates greater pixel-to-pixel variation, but this is not necessarily informative: if the additional spread is driven by noise rather than underlying structure, it can introduce spurious contrast into the line-of-sight prior. The high-weight tail is especially relevant because these pixels contribute more strongly to the host-galaxy probability. Variations to the distribution's symmetry, particularly its tails, are important as they can alter the relative contributions of low- and high-weight pixels. 
Changes in the distribution as a function of redshift are also essential to consider. If observational effects alter the weights differently across redshift bins, they can change the relative prior probability assigned to different parts of the line of sight. This may weaken the redshift information available for $H_0$, or, if the re-weighting is systematic, shift the inferred posterior. Thus, the distributions in Figure~\ref{fig:hist} provide qualitative context for the cosmological results: reduced contrast can diminish the constraining power of the intensity-map prior, while redshift-dependent distortions may affect the inferred $H_0$.
These histograms summarise the distribution of pixel weights but not their spatial arrangement. Thus, maps with similar distributions may yield different LoS priors if their high-weight pixels lie in different regions of the sky or at different redshifts. The impact is event-dependent and is conditioned by their sky localisation and distance posterior. Therefore, we only use these histograms as diagnostic tools, and eventually assess the cosmological consequences of these map-level transformations through the resulting $H_0$ posteriors.

In subsection~\ref{subsec:results_ideal_to_realistic}, we address three main questions: \textbf{(1)} Can redshift priors informed by the intrinsic \hi\ map substantially improve the $H_0$ precision compared to the spectral-siren analysis? \textbf{(2)} To what extent do measurement uncertainties in the GW masses, sky position and luminosity distance estimates propagate into the cosmological constraints? \textbf{(3)} How do the observational challenges associated with 21-cm IM affect the cosmological inference, and which effects are most detrimental to the information extracted from the \hi\ maps? Eventually, we discuss possible mitigations and additional considerations in subsection~\ref{subsec:results_mitigations}, and show the results for the full populations inference in subsection~\ref{subsec:results_full_inference}.

\subsection{From idealised to realistic forecasts} \label{subsec:results_ideal_to_realistic}
The idealised case, assuming observationally unaltered \hi\ maps and neglecting GW parameter uncertainties, shows that including a \hi-informed redshift prior can substantially improve the precision on $H_0$. As shown in Figure~\ref{fig:H0_posteriors_forest}, the $1\sigma$ uncertainty is reduced by \hiNoHiImprovementMS{} for both MeerKLASS-UHF and  SKA-Mid1 compared to the spectral-siren analysis, and allows us to reach sub-percent precision on $H_0$. This highlights the potential of \hi\ intensity maps to provide constraining redshift information along the GW lines of sight.
\newline

However, these results should only be regarded as an upper limit on the constraining power of the method for one year of observations with the LVK network at O5 sensitivity and the complete MeerKLASS-UHF or SKA-Mid1 surveys. Including GW parameter uncertainties results in a major decrease in precision, with the 68\% confidence interval width going from sub-percent precision to just below \hiGwPrecisionMS{} for both surveys (see Figure~\ref{fig:H0_posteriors_forest}). This highlights that GW parameter uncertainties are the main limiting factor for the method when the observed \hi\ map provides highly informative redshift priors. Conversely, when the angular resolution of the map is poorer than the GW sky localisation, GW errors become less dominant in limiting the $H_0$ constraints, although uncertainties in the detected masses and luminosity distances still contribute significantly. However, we note that it still performs better than the spectral-siren case with GW errors: we find that the $1\sigma$ uncertainty decreases by \hiGwNoHiGwImprovementM{} and \hiGwNoHiGwImprovementS{} for MeerKLASS-UHF and SKA-Mid1, respectively. In parallel, including observational effects to produce realistic forecasts reduces the cosmological information carried by the maps, substantially diminishing the relative impact of GW parameter uncertainties. The results suggest that SKA-Mid1 is in general more robust against instrumental limitations due to the enhanced \hi-to-noise ratio, larger sky area, and redshift range of SKAO compared to its predecessor MeerKAT, as illustrated by Figure~\ref{fig:hist}. We discuss these effects individually below.

\begin{enumerate}
    \item \textbf{Beam convolution.} Beam convolution smooths the angular structure of the \hi\ field, suppressing small-scale fluctuations and reducing the contrast between over-dense and under-dense regions. This is clearly visible in the temperature-weight distribution (see Figure~\ref{fig:hist}), which becomes more concentrated around intermediate values compared to that of the original maps, with the high- and low-weight tails suppressed. This effect is also redshift dependent, as the beam size increases towards lower observing frequencies (higher redshifts). 
    As a result, including beam convolution degrades the precision of the $H_0$ constraints by a factor of $\mathcal{O}$(10) relative to the ideal \hi-only case (see Figure~\ref{fig:H0_posteriors_forest}). The effect is qualitatively similar for both surveys, although their different frequency ranges result in different beam sizes (see Table~\ref{tab:survey_specs}).

    \item \textbf{Instrumental noise.} Instrumental noise introduces stochastic fluctuations in the \hi\ field, whose relative importance increases towards higher redshift. Its impact is therefore strongly dependent on the map \hi-to-noise ratio: at the effective redshift, the \hi-to-noise ratio is $0.9$ for MeerKLASS-UHF compared to $1.5$ for SKA-Mid1 (see Table~\ref{tab:survey_specs}). This difference is clearly reflected in the distributions of temperature weights shown in Figure~\ref{fig:hist}. For MeerKLASS-UHF, instrumental noise substantially reshapes the original distribution, reducing its skewness and bringing it much closer to an approximately normal distribution of temperature weights. The intrinsic \hi\ structure is consequently largely obscured, particularly at the high-redshift end, where the noise becomes increasingly important. In contrast, the SKA-Mid1 distribution remains much closer to the original case, although a similar progressive flattening is visible at high redshift. These differences propagate into the cosmological inference results shown in Figure~\ref{fig:H0_posteriors_forest}. For SKA-Mid1, the parameter precision remains unchanged, whereas for MeerKLASS-UHF, instrumental noise noticeably decreases the $H_0$ precision and produces a critical positive bias in both $H_0$ and $\Omega_m$. Indeed, when including instrumental noise, the posterior distribution features two narrow peaks: one centred on the fiducial value and a second at higher $H_0$, well above the Hubble tension interval, i.e. $[67,74] \rm ~km~s^{-1}~Mpc^{-1}$. While the second peak becomes negligible in the SKA-Mid1 scenario, it is instead preferred in the MeerKLASS-UHF case. To investigate the origin of the bias, we fix $\Omega_m$ to its fiducial value. This significantly reduces the bias, although the second peak remains dominant with twice the posterior amplitude of the first one. In particular, the posterior median remains centred on the second peak, while its lower $1\sigma$ uncertainty extends up to the first peak, centred on the fiducial value (see the MeerKLASS-UHF results in Figure~\ref{fig:H0_posteriors_forest}). The uncertainty of the peak at the fiducial value is sub-percent, approximately three times smaller than in the original-map case. These results suggest that the redshift dependence of the noise is particularly important at low \hi-to-noise, where the observational transformation becomes sufficiently strong to modify the effective line-of-sight prior and consequently alter the inferred cosmological parameters.

    \item \textbf{Foreground cleaning.} Foreground cleaning removes spectrally coherent foreground modes, but may also remove cosmological \hi\ fluctuations along with the foreground emission. The effect is marginal for SKA-Mid1, whose temperature-weight distribution remains close to the original one except at $z_{\rm max}$, while the deviation is more pronounced for MeerKLASS-UHF (see Figure~\ref{fig:hist}). In both cases, the high-temperature-weight region is relatively well preserved, while an extended tail towards lower temperature weights develops, indicating that the cleaning procedure modifies the low-contrast part of the distribution more strongly than the regions associated with the largest host probabilities. The overall shape is therefore not dramatically distorted, but the important redshift dependence is problematic for the consistency of the inferred line-of-sight prior. This importantly reduces the $H_0$ precision for MeerKLASS-UHF whereas for SKA-Mid1 the precision is very slightly improved (see Figure~\ref{fig:H0_posteriors_forest}). Note that we adopt different levels of foreground cleaning for the two surveys. A detailed discussion of the impact of foreground cleaning on the resulting maps and cosmological inference is provided in Appendix~\ref{app:foreground}. The difference in performance may be related to the larger frequency range and number of frequency channels available to SKA-Mid1, which provide more spectral information for identifying foreground modes. However, other factors, such as the \hi-to-foreground amplitude ratio and the angular scales retained in the maps, may also contribute. A systematic study of these dependencies is left for future work.

    \item \textbf{Partial sky coverage.} Partial sky coverage reduces the redshift information available to GW events whose localisation regions extend beyond the \hi\ survey footprint. Unlike the instrumental transformations considered above, this effect does not directly alter the temperature-weight distribution of the observed pixels. Instead, it removes part of the spatial structure entering the line-of-sight prior. Its impact therefore depends on the overlap between the GW localisation regions and the survey footprint, as well as on the redshift distribution of the detected events. As reported in Figure~\ref{fig:H0_posteriors_forest}, we find that incomplete sky coverage has a limited impact on the $H_0$ precision compared with the ideal \hi\ case represented by the original map, with the effect being more pronounced for MeerKLASS-UHF which covers a smaller sky area than for SKA-Mid1. This behaviour is consistent with sky coverage acting primarily as a loss of information rather than as a systematic distortion of the tracer. For MeerKLASS-UHF, the precision improves from $1.1\%$ with 10,000 deg$^2$ (MeerKLASS-UHF nominal sky area) to $0.4\%$ with 20,000 deg$^2$ (SKA-Mid1 nominal sky area). Conversely, for SKA-Mid1, the precision degrades from $0.21\%$ for its nominal sky coverage to $0.24\%$ when adopting the MeerKLASS-UHF nominal sky area. These results suggest that the sky coverage of the \hi\ surveys does not translate into a linear improvement in the $H_0$ precision.

    \item \textbf{Combined effect.} As illustrated by Figure~\ref{fig:hist}, the combined effect of beam convolution, instrumental noise, foreground cleaning, and partial sky coverage produces a substantially altered \hi-informed redshift prior in both surveys. For MeerKLASS-UHF, the temperature-weight distribution becomes substantially flatter than all of the other observational scenarios, indicating that the map retains a reduced amount of the redshift-dependent structure that distinguishes the \hi-informed prior from the uninformative homogeneous field case. The combined observational effects therefore largely erase the additional information provided by the \hi\ tracer, substantially degrading the $H_0$ precision relative to the ideal \hi\ case and leaving a \allNoHiImprovementM{} improvement relative to the spectral-siren analysis. For SKA-Mid1, the combined distribution is dominated by the effect of beam smoothing, with noise and foreground cleaning producing additional suppression of the temperature-weight contrast. However, the resulting distribution remains centred relatively close to unity. This reduces the preferential weighting of either over- or under-dense regions and may explain why the combined observational effects primarily degrade the precision rather than resulting in a bias. The resulting $H_0$ precision is considerably decreased relative to the original \hi\ case, while retaining a \allNoHiImprovementS{} improvement relative to the spectral-siren analysis. It is interesting to note that the combined effect of all observational limitations is substantially larger than that of the sum of the individual effects, particularly for SKA-Mid1. When all effects are included, the temperature-weight distribution deviates more strongly from that of the original map for SKA-Mid1 than for MeerKLASS-UHF, resulting in a substantially greater loss of precision in the cosmological inference.
\end{enumerate}

The comparison between MeerKLASS-UHF and SKA-Mid1 depends on both survey properties and how they affect the lines of sight, and GW event selection. When observational effects are considered individually, SKA-Mid1 generally gives tighter constraints, consistent with its broader redshift coverage, higher \hi-to-noise, and wider sky coverage (see Table~\ref{tab:survey_specs}). This is important because the added power of dark-siren analyses depends on the line-of-sight information content: a sharper $\mathbb{P}_{\rm LoS}(z)$ provides stronger leverage on the redshift-distance relation, whereas smoothing, noise, foreground removal, or missing sky coverage progressively reduce this information and may introduce biases at low \hi-to-noise or \hi-to-foreground ratios. In the spectral-siren case, however, SKA-Mid1 yields weaker constraints than MeerKLASS-UHF (see Figure~\ref{fig:H0_posteriors_forest}), while we would expect similar results for independent spectral analysis with a similar number of events. This outcome must arise from event selection: the forecasts use independently drawn GW catalogues, whose differing event properties can affect the inferred constraints. Indeed, the fact that the two realisations also produce slightly different constraints in the spectral-siren case illustrates the realisation-to-realisation variability expected in these forecasts. The final comparison therefore combines these two competing effects and should be interpreted with that caveat in mind, motivating a more extensive robustness study in future work. Using a common GW event catalogue, or analysing multiple catalogue realisations, would help disentangle these contributions.

\subsection{Possible mitigations and additional considerations} \label{subsec:results_mitigations}
Although we find no concerning systematic biases in our fiducial combined results, it is crucial to further understand and investigate the differential impact of observational effects as a function of redshift and mitigate it by either modelling the redshift evolution of the observational effects inside the pipeline or masking low-confidence redshift ranges. 
To begin with, the redshift evolution of the beam size and instrumental-noise fluctuation amplitude is described by Equations~\ref{eq:Gaussian_beam} and \ref{eq:radiometer} and it should be modelled within the pipeline to ensure robustness. 
Then, a promising direction for further modelling these effects would be to quantify their impact on the temperature weights, and subsequently on the redshift priors, from a simulation-based systematic study. In particular, for effects that do not have a simple analytical description, such as foreground removal or the overall combined effect, one could construct a transfer-function-like correction by characterising their impact with a large number of realisations and different \hi\ simulation assumptions. This could then be used to calibrate the observed temperature-weight distribution, and ultimately the line-of-sight redshift prior, against the original \hi\ field. 

In connection with this point, a relevant caveat is that the present analysis does not model the full redshift evolution of the underlying \hi\ tracer: in particular, the mean brightness temperature $\overline{T}_{\rm H_I}$ is constant in our current simulations. However, note that we include a dependence on the functional form of the SFR, which effectively models the redshift dependence of the GW and \hi\ distributions, i.e. $\overline{T}_{\rm H_I}(z) \propto \overline{T}_{\rm H_I}\,\Psi(z)$. This was done to ensure consistency with the GW population observed by LVK. We leave consideration of other parameterisations to future work.  Moreover, we assume that the value of $\overline{T}_{\rm H_I}$ is known, whereas in reality, it would be interesting to jointly inferred it or simply measured it from the \hi\ field. This is potentially important as we use this value in our temperature-weight expression and because our results suggest that the interplay between the evolution of the underlying \hi\ field and that of observational effects may be complex and difficult to separate. If not properly modelled, this interplay could introduce biases in the inferred cosmological parameters due to distorted line-of-sight redshift priors. This motivates the treatment of these observational effects as independent nuisance parameters and their marginalisation in the analysis.  A study including a realistic redshift evolution of $\overline{T}_{\rm H_I}$ is left for future work. Meanwhile, measuring $\overline{T}_{\rm H_I}$ directly from the realistic map may help mitigate the impact of these observational effects, as we find that the median of the temperature-weight distribution varies substantially between scenarios. Accounting for this variation when constructing the redshift prior could therefore provide a more accurate description of the information retained in the observed map. 
Similarly, the \hi\ bias $b_{\rm HI}(k,z)$ is not constant at nonlinear scales and this should be accounted for when using lower redshift or higher map  resolution.

\begin{figure*}
    \centering
    \includegraphics[width=\linewidth]{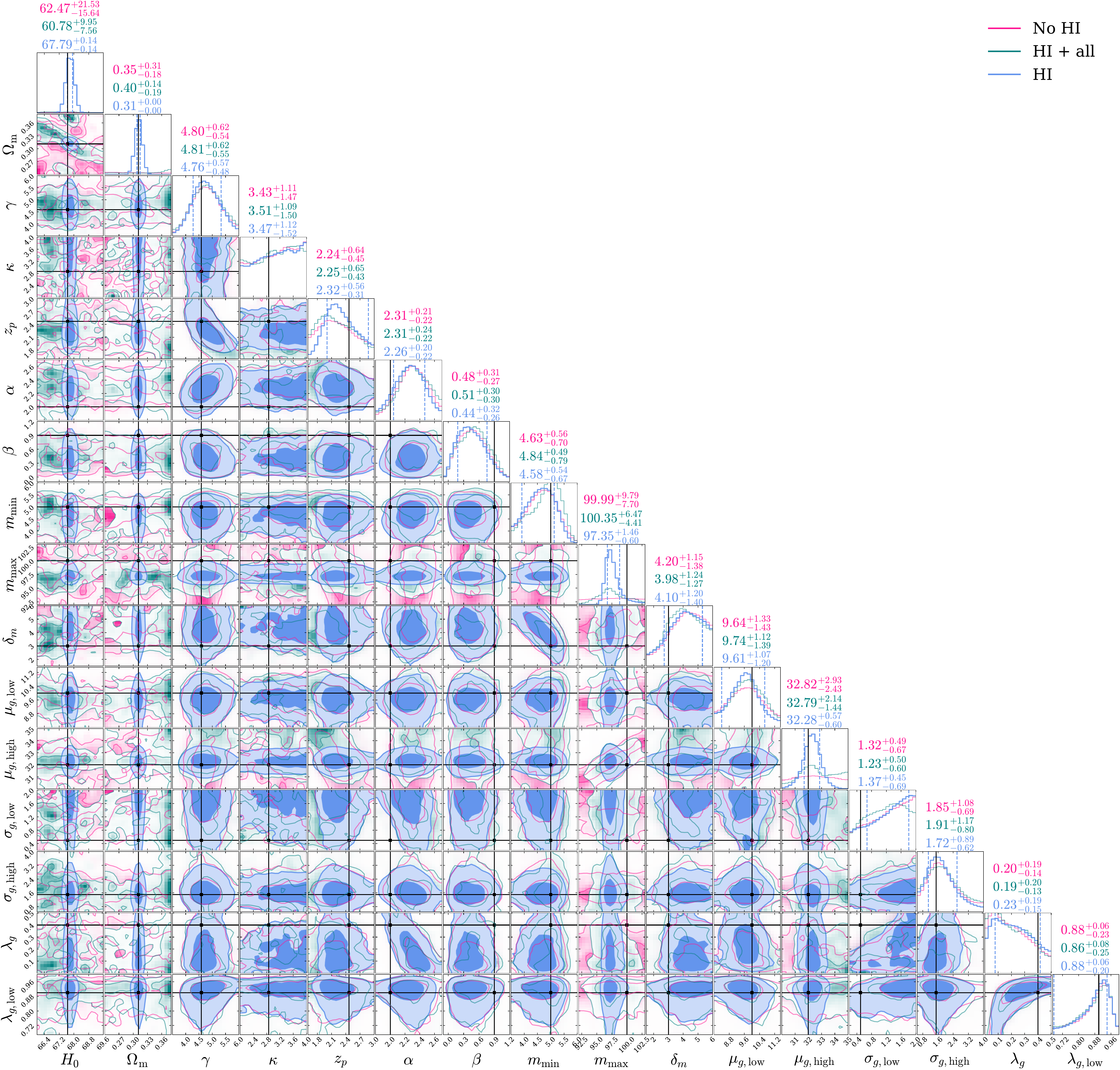}
    \caption{Posterior distributions for the cosmological and population hyper-parameters inferred in the SKA-Mid1 scenario. We compare results obtained without \hi\ map information to those obtained with SKA-Mid1 map support, considering both an original \hi\ map and a realistic map including observational effects: beam convolution, instrumental noise, incomplete sky coverage, and foreground removal. The measurement uncertainties in the GW parameters are neglected. The shaded contours show the 1$\sigma$- and 2$\sigma$-equivalent credible regions, while the solid lines indicate the fiducial values of the hyper-parameters. Prior ranges and fiducial values are given in Table~\ref{tab:hyper-parameters}). $H_0$ has units of km~s$^{-1}$~Mpc$^{-1}$, while $m_{\rm min}$, $m_{\rm max}$, $\delta_{\rm m}$, $\mu_{\rm g, low}$, $\mu_{\rm g, high}$, $\sigma_{\rm g, low}$, and $\sigma_{\rm g, high}$ are expressed in solar masses.}
    \label{fig:corner_plot}
\end{figure*}
\subsection{A joint population and cosmological inference} \label{subsec:results_full_inference}
Finally, it is informative to examine which parameters drive the cosmological constraints and how the inclusion of the \hi\ map affects the population inference as a whole. Figure~\ref{fig:corner_plot} shows the posterior distributions of the cosmological and population hyper-parameters for ideal and realistic \hi\ scenarios, as well as the standard-sirens case. The strongest correlation involving the cosmological parameters is the $H_0$--$\Omega_m$ degeneracy, which is broad in the absence of \hi\ information and becomes tightly constrained in the ideal \hi\ case, represented by the original map. This demonstrates that the \hi\ map provides information that is directly relevant to the distance--redshift relation of the GW sources. In contrast, correlations between the cosmological and population hyper-parameters are generally weaker, indicating that the additional constraining power does not arise from a direct tightening of the population model.
Indeed, most population parameters are remarkably stable across the three scenarios, with the inferred values generally consistent within their $68\%$ credible regions and with comparable uncertainties. The redshift-distribution parameters, in particular, show little variation between the analyses, suggesting that the \hi\ information primarily improves the redshift inference of individual GW sources rather than substantially modifying the inferred global merger-rate evolution. An important exception is the upper mass cut-off, $m_{\rm max}$, which is substantially better constrained when \hi\ information is included, with its uncertainty decreasing from approximately $9\%$ in the spectral case to less than $2\%$ for the original \hi\ scenario. This improvement can be understood from the improved redshift information provided by the \hi\ map: the conversion from detector-frame to source-frame masses depends directly on redshift, allowing features in the intrinsic mass distribution, such as the upper mass cut-off, to be more precisely located. Similar results are observed for the localised mass features such as $\mu_{\rm g, high}$, $\mu_{\rm g, low}$, and  $m_{\rm min}$. The fact that the main changes induced by the \hi\ map occur in $H_0$, $\Omega_m$, while most population parameters remain largely unchanged, provides further evidence that the \hi\ map acts primarily by improving the redshift information associated with the GW events. Once realistic observational effects are included, the $H_0$--$\Omega_m$ posterior becomes substantially broader and shifts away from the ideal solution, while the population parameters remain comparatively stable. This suggests that the dominant impact of the observational limitations is to distort the \hi-induced line-of-sight redshift prior, rather than to fundamentally alter the inferred BBH population.
\newline

\section{Conclusion and discussion}
\label{sec:conclusion}
In this work, we have investigated the potential of \hi\ IM to improve cosmological inference from GW dark sirens. By incorporating the \hi\ intensity field into a hierarchical Bayesian framework, we use the redshift-dependent structure of the tracer to construct an informative line-of-sight prior for the redshift of BBH sources. We consider two representative surveys, MeerKLASS-UHF and SKA-Mid1, and progressively incorporate the main observational effects affecting real intensity-mapping observations, namely beam convolution, instrumental noise, incomplete sky coverage, and foreground cleaning. We forecast one year of observations at O5 sensitivity using 146 (MeerKLASS-UHF) and 147 (SKA-Mid1) dark sirens with $\mathrm{SNR}>12$.

Our results show that the inclusion of idealised \hi\ maps can substantially improve the cosmological inference relative to the spectral-siren case, yielding a sub-percent measurement of $H_0$ and a \hiNoHiImprovementMS{} reduction in its uncertainty. This idealised scenario therefore represents an upper limit to the constraining power achievable with perfect \hi\ maps and perfectly measured GWs. When beam convolution, instrumental noise, partial sky coverage, and foreground cleaning are incorporated, the improvement is reduced to \allNoHiImprovementM{} for MeerKLASS-UHF and \allNoHiImprovementS{} for SKA-Mid1. The individual impact of these observational limitations is moderate, particularly for SKA-Mid1, although they can introduce multimodality in the posterior distributions. However their combined effect constitutes the dominant limitation in the realistic analysis. We briefly discuss this combined effect at the map level, focusing on the redshift dependence of the observational limitations, and consider potential mitigation strategies to improve the robustness and precision of the analysis.  In addition, GW parameter measurement uncertainties substantially degrade the constraints for idealised \hi\ maps, increasing the uncertainty on $H_0$ from sub-percent precision to below \hiGwPrecisionMS{}. In contrast, their impact is less significant when realistic \hi\ maps are used, with the final constraints reaching \allGwPrecisionM{} and \allGwPrecisionS{} precision for MeerKLASS-UHF and SKA-Mid1, respectively. With our current methodology, the discrepancy between the forecasted precision for two surveys is driven by realisation-to-realisation variability rather than survey specifications. These results demonstrate that \hi\ IM can provide significant additional information on $H_0$ and serve as a complementary probe for GW cosmology, while highlighting the importance of jointly accounting for observational effects in the \hi\ maps and uncertainties in the GW parameter measurements.

For comparison, current LVK results achieve approximately $20\%$ precision on $H_0$ using 235 dark sirens and the Dark Energy Survey Year~6 galaxy catalogue \citep{GWTC5_cosmo}. Our forecasts achieve tighter constraints despite using fewer events, which may be partly due to the favourable overlap between the GW population and \hi\ redshift distributions, and the large sky coverage of the IM surveys

A joint multi-tracer analysis combining \hi\ intensity maps, galaxy catalogues, or other large-scale-structure tracers could provide a more robust redshift prior than any single tracer alone. Combining tracers with complementary redshift ranges, precision, angular resolution, and noise properties would allow us to reduce systematics tied to individual-tracer assumptions, thus improving robustness against observational limitations, and maximising the precision and accuracy of the redshift measurements, similarly to the multi-messenger observation of the binary neutron star GW170817~\citep{GW170817_MM,GW170817_H0}. SKA-Mid1 will have particularly good overlap with DESI~\citep{DESI}, Euclid~\citep{Euclid}, and LSST~\citep{LSST}. In this regard, \cite{Zazzera_2026} explores the triple cross-correlation of ET-like GWs, SKA-Mid1-like intensity maps and LSST-like photometric galaxy surveys in the context of constraining the magnification lensing bias. The results show that such a multi-tracer analysis leads to tighter constraints with respect to the $\rm GW \times IM$ and $\rm GW \times galaxies$ precision. Similarly, \cite{Scelfo_2023} find that cross-correlations of GWs, intensity maps, and resolved galaxies yield competitive constraints for weak lensing and angular clustering measurements. Besides galaxies, galaxy clusters can be a good probe of large-scale structure to explore in multi-tracer frameworks. The study by \cite{Beirnaert_2025} shows an improvement by up to 38\% with respect to the GWTC-3 results with galaxy survey \citep{GWTC3_cosmo}. Alternatively, we could imagine a multi-survey analysis by combining SKAO with other wide 21-cm IM surveys in operation such as CHIME~\citep{CHIME} and FAST~\citep{FAST} in the Northern hemisphere, or HIRAX~\citep{HIRAX}, under development in the Southern Hemisphere.

Furthermore, another way to improve robustness against observational effects is to go beyond the use of the \hi\ field as a redshift prior only and integrate complementary observables in our likelihood expression (Equation~\ref{eq:likelihood}). In particular, the GW--\hi\ cross-power spectrum could exploit the spatial correlation between GW sources and the underlying matter field, providing complementary clustering information. Furthermore, the baryon acoustic oscillation (BAO) scale, measured by SKAO and other surveys, provides an independent standard-ruler probe of the expansion history which can help break the $H_0$--$\Omega_m$ degeneracy as demonstrated by analyses combining BAO with GW observations and CMB information \citep{Afroz_2024,Belgacem_2018,Pierra_2024}.

Most importantly, our results highlight the importance of well-measured, and in particular well-localised, GW detections. Improved detector sensitivity will allow us to select higher-SNR sources with substantially better distance and sky-localisation measurements while maintaining or increasing the number of useful events. This will become particularly relevant with third-generation detectors such as the Einstein Telescope (ET) and Cosmic Explorer (CE), whose sensitivities will greatly exceed those expected for the LVK network during O5~\citep{Punturo_2010_ET,Reitze_2019_CE}. A forecast of the synergy between SKA-Mid and ET for dark-siren cosmology is presented in \cite{Dupletsa_2026}. Using $1500$ detections with $\mathrm{SNR}>150$ from a single ET detector, $H_0$ can be constrained with 8\% precision, without relying on spectral information. Since sky localisation is particularly important for exploiting the angular structure of the \hi\ field, the constraints would further improve when using ET in a network with CE.

\section*{Acknowledgements}

The authors would like to thank Monica Seglar-Arroyo and the members of the GWCats consortium for useful discussions.

This work is partially supported by the Spanish MCIN/AEI/10.13039/501100011033 under the Grants No. PID2020-113701GB-I00 and PID2023-146517NB-I00, some of which include ERDF funds from the European Union, and by the MICINN with funding from the European Union NextGenerationEU (PRTR-C17.I1) and by the Generalitat de Catalunya. IFAE is partially funded by the CERCA programme of the Generalitat de Catalunya.
This project has received funding from the European Union’s Horizon Europe research and innovation programme under the Marie Skłodowska-Curie grant agreement No. 10181337.
The research leading to these results has received funding from the European Union under grant agreement N.~101299291 (GWCats).
SC is funded by a UKRI Stephen Hawking Fellowship (grant reference EP/U536751/1). 
TB acknowledges support from Royal Society grant number URF\textbackslash R\textbackslash 231006 and ERC Starting Grant SHADE (grant no.\,StG 949572).
SM is supported by ERC grant GravitySirens 101163912. Funded by the European Union.
M. Spinelli is supported by the French government through the France 2030 investment plan managed by the National Research Agency(ANR), as part of the Initiative of Excellence Université Côte d’Azur under reference number ANR-15-IDEX-01 and by the French Programme National de Cosmologie et Galaxies (PNCG project CIMES).

The data production, processing, and analysis tools for this paper have been developed, implemented and operated in collaboration with the Port d’Informació Científica (PIC) data centre. PIC is maintained through a collaboration agreement between the Institut de Física d’Altes Energies (IFAE) and the Centro de Investigaciones Energéticas, Medioambientales y Tecnológicas (CIEMAT). The authors are grateful for computational resources provided by the LIGO Laboratory and supported by National Science Foundation Grants PHY-0757058 and PHY-0823459. This material is based upon work supported by NSF's LIGO Laboratory which is a major facility fully funded by the National Science Foundation.

The first five authors are listed in order of contribution. DNL led the study and drafted the manuscript, developing the framework in collaboration with the co-authors, and carrying out the simulations and analysis. KL and TB contributed their expertise in GW cosmology, with KL sharing a previously developed GW simulation pipeline that was adapted for this study. SC and IH contributed their expertise in 21-cm IM with the MeerKLASS and SKA surveys, with SC sharing a previously developed intensity-map simulation pipeline that was adapted for this study. The remaining authors participated in the broader collaborative effort to explore synergies between 21-cm IM and GW dark sirens and are listed alphabetically.

\section*{Data Availability}

The data used in this work are original and were generated as part of this study. The datasets, including simulated maps and mock GW catalogues, are available from the corresponding authors upon reasonable request. Additional details on the data generation process and analysis pipelines can be provided to facilitate reproducibility.



\bibliographystyle{mnras}
\bibliography{references}

@article{Schutz_1986,
author = "Schutz, B.",
title = "Determining the Hubble constant from gravitational wave observations",
journal = "\nat",
volume = "323",
pages = "310-311",
year = "1986",
doi = {10.1038/323310a0}
}

@article{Planck2020,
   title={<i>Planck</i>2018 results: VI. Cosmological parameters},
   volume={641},
   ISSN={1432-0746},
   url={http://dx.doi.org/10.1051/0004-6361/201833910},
   DOI={10.1051/0004-6361/201833910},
   journal={\aap},
   publisher={EDP Sciences},
   author={{Planck Collaboration}},
   year={2020},
   month=Sept, pages={A6} }

@article{Riess2022,
author = "Riess, A. G. and others",
title = {A Comprehensive Measurement of the Local Value of the Hubble Constant with 1 km s\textsuperscript{-1} Mpc\textsuperscript{-1} Uncertainty from the Hubble Space Telescope and the SH0ES Team},
journal = "\apj",
volume = "934",
number = "1",
pages = "L7",
year = "2022",
doi = {10.3847/2041-8213/ac5c5b}
}

@article{GW170817_MM,
author = "{The LIGO Scientific Collaboration and the Virgo Collaboration} and others",
title = "Multi-messenger Observations of a Binary Neutron Star Merger",
journal = "\apjl",
volume = "848",
pages = "L12",
year = "2017",
doi = {10.3847/2041-8213/aa91c9}
}

@article{GW170817_detection,
author = "{The LIGO Scientific Collaboration and the Virgo Collaboration}",
title = "GW170817: Observation of Gravitational Waves from a Binary Neutron Star Inspiral",
journal = "\prl",
volume = "119",
pages = "161101",
year = "2017",
doi = {10.1103/PhysRevLett.119.161101}
}

@article{GW170817_H0,
author = {{The LIGO Scientific Collaboration and the Virgo Collaboration} and others},
title = "A gravitational-wave standard siren measurement of the Hubble constant",
journal = "\nat",
volume = "551",
pages = "85--88",
year = "2017",
doi = {10.1038/nature24471},
}

@article{Soares-Santos_2019,
author = "Soares-Santos, M. and others",
title = "First Measurement of the Hubble Constant from a Dark Standard Siren using the Dark Energy Survey Galaxies and the LIGO/Virgo Binary-Black-hole Merger GW170814",
journal = "\apjl",
volume = "876",
pages = "L7",
year = "2019",
doi = {10.3847/2041-8213/ab14f1}
}

@article{Palmese_2020,
year = {2020},
month = {sep},
volume = {900},
number = {2},
pages = {L33},
author = {Palmese, A. and others},
title = {A Statistical Standard Siren Measurement of the Hubble Constant from the LIGO/Virgo Gravitational Wave Compact Object Merger GW190814 and Dark Energy Survey Galaxies},
journal = {\apjl},
doi = {10.3847/2041-8213/abaeff},
}

@article{GWTC4_cat,
doi = {10.3847/2041-8213/ae2c74},
year = {2026},
month = {jun},
volume = {1004},
number = {2},
pages = {L22},
author = {{The LIGO Scientific Collaboration, the Virgo Collaboration, and the KAGRA Collaboration}},
title = {GWTC-4.0: Updating the Gravitational-wave Transient Catalog with Observations from the First Part of the Fourth LIGO–Virgo–KAGRA Observing Run},
journal = {\apjl},
}

@article{Del_Pozzo_2012,
author = {Del Pozzo, W.},
title = {Inference of cosmological parameters from gravitational waves: Applications to second generation interferometers},
journal = {\prd},
volume = {86},
pages = {043011},
year = {2012},
doi = {10.1103/PhysRevD.86.043011}
}

@article{Fishbach_2019,
doi = {10.3847/2041-8213/aaf96e},
year = {2019},
month = {jan},
volume = {871},
number = {1},
pages = {L13},
author = {Fishbach, M. and others},
title = {A Standard Siren Measurement of the Hubble Constant from GW170817 without the Electromagnetic Counterpart},
journal = {\apjl},
}

@article{GWTC3_cosmo,
author = "{The LIGO Scientific Collaboration and the Virgo Collaboration}",
title = "Constraints on the Cosmic Expansion History from GWTC–3", 
journal = "\apj",
volume = "949",
pages = "76",
year = "2023",
doi = {10.3847/1538-4357/ac74bb},
}

@misc{GWTC5_cosmo,
  title={GWTC-5.0: Constraints on the Cosmic Expansion Rate and Modified Gravitational-wave Propagation}, 
  author={{The LIGO Scientific Collaboration and the Virgo Collaboration and the KAGRA Collaboration}},
  year={2026},
  eprint={2605.27227},
  archivePrefix={arXiv},
  primaryClass={astro-ph.CO},
}

@article{Gray_2020,
author = "Gray, R. and others",
title = {Cosmological inference using gravitational wave standard sirens: A mock data analysis},
journal = "\prd",
volume = "101",
pages = "122001",
year = "2020",
doi = {10.1103/PhysRevD.101.122001}
}

@article{Gray_2022,
    author = {Gray, R and Messenger, C and Veitch, J},
    title = {A pixelated approach to galaxy catalogue incompleteness: improving the dark siren measurement of the Hubble constant},
    journal = {\mnras},
    volume = {512},
    number = {1},
    pages = {1127-1140},
    year = {2022},
    month = {05},
    doi = {10.1093/mnras/stac366},
}

@article{Gray_2023,
author = "Gray, R. and others",
title = {Joint Cosmological and Gravitational-Wave Population Inference Using Dark Sirens and Galaxy Catalogues},
journal = "\jcap",
volume = "12",
pages = "023",
year = "2023",
doi = {10.1088/1475-7516/2023/12/023}
}

@article{Mastrogiovanni_2023,
author = "Mastrogiovanni, S. and others",
title = {Joint population and cosmological properties inference with gravitational waves standard sirens and galaxy surveys},
journal = "\prd",
volume = "108",
pages = "042002",
year = "2023",
doi = {10.1103/PhysRevD.108.042002}
}

@article{Villaescusa-Navarro_2018,
author = "Villaescusa-Navarro, F. and others",
title = "Ingredients for 21 cm Intensity Mapping",
journal = "\apj",
volume = "866",
pages = "135",
year = "2018",
doi = {10.3847/1538-4357/aadba0}
}

@article{Bernal_2022,
  author  = {Bernal, Jos{\'e} Luis and Kovetz, Ely D.},
  title   = {Line-intensity mapping: theory review with a focus on star-formation lines},
  journal = {\aapr},
  volume  = {30},
  number  = {1},
  pages   = {5},
  year    = {2022},
  doi     = {10.1007/s00159-022-00143-0},
  archivePrefix = {arXiv},
  eprint  = {2206.15377}
}

@misc{Kovetz_2017,
author = "Kovetz, E. D. and others",
title = {Line Intensity Mapping: 2017 Status Report},
journal = "Submitted to \physrep",
year = "2017",
eprint = {1709.09066},
archivePrefix = {arXiv},
primaryClass = {astro-ph.CO}
}

@article{Scelfo_2022,
author = "Scelfo, G. and others",
title = {Gravitational waves $\times$ HI intensity mapping: cosmological and astrophysical applications},
journal = "\jcap",
volume = "2022",
pages = "004",
year = "2022",
doi = {10.1088/1475-7516/2022/01/004}
}

@misc{Schulz_2026,
  title={Cosmology beyond standard sirens: cross-correlation of gravitational waves and neutral hydrogen intensity mapping}, 
  author={M. Schulz and others},
  year={2026},
  eprint={2607.26193},
  archivePrefix={arXiv},
  primaryClass={astro-ph.CO},
}

@misc{SKAObook,
  title={Cosmology from Synergies Between SKAO Surveys and Gravitational Wave Observations}, 
  author={Tessa Baker and others},
  year={2026},
  eprint={2606.26011},
  archivePrefix={arXiv},
  primaryClass={astro-ph.CO},
}

@article{Mukherjee_2022,
author = "Mukherjee, S. and Dizgah, A. M.",
title = "Cross-correlating Line Intensity Maps with Binary Black Hole Mergers to Constrain the Delay Time Distribution",
journal = "\apjl",
volume = "937",
pages = "L27",
year = "2022",
doi = {10.3847/2041-8213/ac903b},
}

@misc{Dupletsa_2026,
  author = "Dupletsa, U. and others",
  title={Radio sirens: inferring $H_0$ with binary black holes and neutral hydrogen in the era of the Einstein Telescope and the SKA Observatory}, 
  year={2026},
  eprint={2605.12606},
  archivePrefix={arXiv},
  primaryClass={astro-ph.CO},
}

@misc{Santos_2017,
author = "Santos, M. G. and others",
title = {MeerKLASS: MeerKAT Large Area Synoptic Survey},
journal = "MeerKAT Science: On the Pathway to the SKA, Stellenbosch, 25-27 May 2016",
year = "2017",
eprint = "1709.06099",
archivePrefix = "arXiv",
primaryClass = "astro-ph.CO",
}

@article{Battye_2013,
author="Battye, R. A. and others",
title = {H I intensity mapping: a single dish approach},
journal = "\mnras",
volume={434},
number={2},
pages={1239–1256},
DOI={10.1093/mnras/stt1082},
year={2013}, 
}

@article{gridimp,
author = "Cunnington, S. and Wolz, L.",
title = {Accurate Fourier-space statistics for line intensity mapping: Cartesian grid sampling without aliased power},
journal = "\mnras",
volume = "528",
pages = "5586–5600",
year = "2024",
doi = "10.1093/mnras/stae333",
}

@article{healpix,
author = "Gorski, K. M. and others",
title = {HEALPix: A Framework for High Resolution Discretization and Fast Analysis of Data Distributed on the Sphere},
journal = "\apj",
volume = "622",
pages = "759-771",
year = "2005",
doi = "10.1086/427976"
}

@article{bilby,
author = "Ashton, G. and others",
title = {BILBY: A User-friendly Bayesian Inference Library for Gravitational-wave Astronomy},
journal = "\apjs",
volume = "241",
pages = "27",
year = "2019",
doi = "10.3847/1538-4365/ab06fc"
}

@article{MeerKLASS_Lband_2025,
    author = "{MeerKLASS Collaboration}",
    title = {MeerKLASS L-band deep-field intensity maps: entering the HI dominated regime},
    journal = {\mnras},
    volume = {537},
    number = {4},
    pages = {3632-3661},
    year = {2025},
    month = {03},
    doi = {10.1093/mnras/staf195},
}

@article{Dicke_1946,
author = "Dicke, R. H.",
journal = "Rev. Sci. Instrum.",
title = "The Measurement of Thermal Radiation at Microwave Frequencies",
volume = "17",
pages = "268--275",
year = "1946",
doi = "10.1063/1.1770483"
}

@inproceedings{Campbell_2002,
author = "Campbell, D. B.",
title = "Single-Dish Radio Astronomy: Techniques and Applications",
booktitle = "\aspc",
volume = "278",
pages = "81--90",
year = "2002",
editor = "Stanimirović, S. and Altschuler, D. and Goldsmith, P. and Salter, C.",
series = {\aspc},
}

@article{SKA_Cosmology_SWG_2020, 
author = "{SKA Cosmology SWG}",
title = {Cosmology with Phase 1 of the Square Kilometre Array; Red Book 2018: Technical specifications and performance forecasts},
journal = "\pasa",
volume = "37",
pages = "e007",
year = "2020",
doi = "10.1017/pasa.2019.51",
}

@article{Cunnington_2022,
   title={Detecting the power spectrum turnover with HI
                    intensity mapping},
   volume={512},
   ISSN={1365-2966},
   DOI={10.1093/mnras/stac576},
   number={2},
   journal={\mnras},
   publisher={Oxford University Press (OUP)},
   author={Cunnington, Steven},
   year={2022},
   month=Mar, pages={2408–2425} }

@article{Cunnington_2021,
author = "Cunnington, S. and others",
title = {21cm foregrounds and polarization leakage: a user's guide on cleaning and mitigation strategies},
journal = "\mnras",
volume = "504",
pages = "208-227",
year = "2021",
doi = "10.1093/mnras/stab856",
}

@article{Spinelli_2022,
    author = {Spinelli, Marta and others},
    title = {SKAO HI intensity mapping: blind foreground subtraction challenge},
    journal = {\mnras},
    volume = {509},
    number = {2},
    pages = {2048-2074},
    year = {2022},
    month = {01},
    issn = {0035-8711},
    doi = {10.1093/mnras/stab3064},
    eprint = {https://academic.oup.com/mnras/article-pdf/509/2/2048/41170415/stab3064.pdf},
}

@article{Madau-Dickinson_2014,
author = "Madau, Piero and Dickinson, Mark",
title = {Cosmic Star-Formation History},
journal = "\araa",
number={1},
pages={415–486},
year = "2014",
doi = "10.1146/annurev-astro-081811-125615"
}

@article{Mandel_2019,
author = "Mandel, Ilya and Farr, Will M and Gair, Jonathan R",
title = {Extracting distribution parameters from multiple uncertain observations with selection biases},
journal = "\mnras",
volume = "486",
pages = "1086-1093",
year = "2019",
doi = "10.1093/mnras/stz896"
}

@article{Vitale_2021,
author = "Vitale, S. and others",
title = {Inferring the Properties of a Population of Compact Binaries in Presence of Selection Effects},
journal = "Springer Singapore",
volume = "Handbook of Gravitational Wave Astronomy",
pages = "1-60",
year = "2021",
doi = "10.1007/978-981-15-4702-7_45-1"
}

@article{icarogw,
author = "Mastrogiovanni, S. and others",
title = {ICAROGW: A python package for inference of astrophysical population properties of noisy, heterogeneous, and incomplete observations},
journal = "\aap",
volume = "682",
pages = "A167",
year = "2024",
doi = "10.1051/0004-6361/202347007"
}

@article{Bharadwaj:2000av,
    author = "Bharadwaj, Somnath and others",
    title = "{Using HI to probe large scale structures at z \textasciitilde\ 3}",
    eprint = "astro-ph/0003200",
    archivePrefix = "arXiv",
    doi = "10.1007/BF02933588",
    journal = "\japa",
    volume = "22",
    pages = "21",
    year = "2001"
}

@article{Battye:2004re,
    author = "Battye, Richard A. and Davies, Rod D. and Weller, Jochen",
    title = "{Neutral hydrogen surveys for high redshift galaxy clusters and proto-clusters}",
    eprint = "astro-ph/0401340",
    archivePrefix = "arXiv",
    doi = "10.1111/j.1365-2966.2004.08416.x",
        journal = "\mnras",
    volume = "355",
    pages = "1339--1347",
    year = "2004"
}

@article{Chang:2007xk,
    author = "Chang, Tzu-Ching and others",
    title = "{Baryon Acoustic Oscillation Intensity Mapping as a Test of Dark Energy}",
    eprint = "0709.3672",
    archivePrefix = "arXiv",
    primaryClass = "astro-ph",
    doi = "10.1103/PhysRevLett.100.091303",
    journal = "\prl",
    volume = "100",
    pages = "091303",
    year = "2008"
}

@article{Wyithe:2007rq,
    author = "Wyithe, Stuart and Loeb, Abraham and Geil, Paul",
    title = "{Baryonic Acoustic Oscillations in 21cm Emission: A Probe of Dark Energy out to High Redshifts}",
    eprint = "0709.2955",
    archivePrefix = "arXiv",
    primaryClass = "astro-ph",
    doi = "10.1111/j.1365-2966.2007.12631.x",
    journal = "\mnras",
    volume = "383",
    pages = "1195",
    year = "2008"
}

@article{Liu:2019awk,
    author = "Liu, Adrian and Shaw, J. Richard",
    title = "{Data Analysis for Precision 21 cm Cosmology}",
    eprint = "1907.08211",
    archivePrefix = "arXiv",
    primaryClass = "astro-ph.IM",
    doi = "10.1088/1538-3873/ab5bfd",
    journal = "\pasp",
    volume = "132",
    number = "1012",
    pages = "062001",
    year = "2020"
}

@article{Cunnington:2025sdr,
    author = "Cunnington, Steven and others",
    title = "{Revealing cosmological fluctuations in 21 cm intensity maps with MeerKLASS: from maps to power spectra}",
    eprint = "2510.27549",
    archivePrefix = "arXiv",
    primaryClass = "astro-ph.CO",
    doi = "10.1007/s10509-026-04547-7",
    journal = "Astrophys. Space Sci.",
    volume = "371",
    number = "2",
    pages = "16",
    year = "2026"
}

@ARTICLE{Foreman-Mackey_2013,
       author = {{Foreman-Mackey}, Daniel and others},
        title = "{emcee: The MCMC Hammer}",
      journal = {\pasp},
         year = 2013,
        month = mar,
       volume = {125},
       number = {925},
        pages = {306},
          doi = {10.1086/670067},
archivePrefix = {arXiv},
       eprint = {1202.3665},
 primaryClass = {astro-ph.IM},
}

@article{Pratten_2020,
   title={Setting the cornerstone for a family of models for gravitational waves from compact binaries: The dominant harmonic for nonprecessing quasicircular black holes},
   volume={102},
   DOI={10.1103/physrevd.102.064001},
   number={6},
   journal={\prd},
   author={Pratten, Geraint and others},
   year={2020},
   month={Sept}, 
}

@article{Zazzera_2026,
   title={Exploring future synergies for large-scale structure between gravitational waves and radio sources},
   volume={547},
   DOI={10.1093/mnras/stag307},
   number={1},
   journal={\mnras},
   author={Zazzera, Stefano and others},
   year={2026},
   month=Feb }

@article{Scelfo_2023,
   title={Testing gravity with gravitational waves × electromagnetic probes cross-correlations},
   volume={2023},
   DOI={10.1088/1475-7516/2023/02/010},
   number={02},
   journal={\jcap},
   author={Scelfo, Giulio and others},
   year={2023},
   month=Feb, pages={010} }

@article{Beirnaert_2025,
   title={A Hubble constant estimation with dark standard sirens and galaxy cluster catalogues},
   volume={542},
   DOI={10.1093/mnras/staf1432},
   number={4},
   journal={\mnras},
   author={Beirnaert, Freija and Dálya, Gergely and Ghosh, Archisman},
   year={2025},
   month=Aug, pages={3346–3353} }

@article{Afroz_2024,
    author = {Afroz, Samsuzzaman and Mukherjee, Suvodip},
    title = {Prospect of precision cosmology and testing general relativity using binary black holes – galaxies cross-correlation},
    journal = {\mnras},
    volume = {534},
    number = {2},
    pages = {1283-1298},
    year = {2024},
    month = {10},
    doi = {10.1093/mnras/stae2139},
}

@misc{Pierra_2024,
      title={Non-Parametric Reconstruction of the Hubble Parameter from the Fourth Gravitational Wave Transient Catalog and DESI Baryonic Acoustic Oscillations}, 
      author={Grégoire Pierra and Alberto Colombo and Simone Mastrogiovanni},
      year={2025},
      eprint={2511.11795},
      archivePrefix={arXiv},
      primaryClass={astro-ph.CO},
}

@article{Belgacem_2018,
   title={Modified gravitational-wave propagation and standard sirens},
   volume={98},
   DOI={10.1103/physrevd.98.023510},
   number={2},
   journal={\prd},
   author={Belgacem, Enis and others},
   year={2018},
   month=July }

@article{CHIME,
doi = {10.3847/1538-4365/ac6fd9},
year = {2022},
month = {jul},
volume = {261},
number = {2},
pages = {29},
author = {{The CHIME Collaboration}},
title = {An Overview of CHIME, the Canadian Hydrogen Intensity Mapping Experiment},
journal = {\apjs},
}

@inproceedings{HIRAX,
   title={HIRAX: a probe of dark energy and radio transients},
   volume={9906},
   doi = {10.1117/12.2234286},
   booktitle={Ground-based and Airborne Telescopes VI},
   editor = {Helen J. Hall and Roberto Gilmozzi and Heather K. Marshall},
   organization = {International Society for Optics and Photonics},
   publisher={SPIE},
   author={Newburgh, L. B. and others},
   year={2016},
   month=Aug, pages={99065X} ,
   series=""
}

@article{FAST,
   title={THE FIVE-HUNDRED-METER APERTURE SPHERICAL RADIO TELESCOPE (FAST) PROJECT},
   volume={20},
   DOI={10.1142/s0218271811019335},
   number={06},
   journal={Int. J. Mod. Phys. D},
   author={Nan, Rendong and others},
   year={2011},
   month=June, pages={989–1024} 
}

@article{Punturo_2010_ET,
doi = {10.1088/0264-9381/27/19/194002},
year = {2010},
month = {sep},
publisher = {},
volume = {27},
number = {19},
pages = {194002},
author = {Punturo, M and others},
title = {The Einstein Telescope: a third-generation gravitational wave observatory},
journal = {Class. Quantum Grav.}
}

@misc{Reitze_2019_CE,
      title={Cosmic Explorer: The U.S. Contribution to Gravitational-Wave Astronomy beyond LIGO}, 
      author={David Reitze and others},
      year={2019},
      eprint={1907.04833},
      archivePrefix={arXiv},
      primaryClass={astro-ph.IM},
}

@misc{DESI,
      title={The DESI Experiment Part I: Science,Targeting, and Survey Design}, 
      author={{DESI Collaboration}},
      year={2016},
      eprint={1611.00036},
      archivePrefix={arXiv},
      primaryClass={astro-ph.IM},
}

@article{Euclid,
	author = {{Euclid Collaboration}},
	title = {Euclid - I. Overview of the Euclid mission},
	DOI= "10.1051/0004-6361/202450810",
	journal = {\aap},
	year = 2025,
	volume = 697,
	pages = "A1",
}

@misc{LSST,
      title={LSST Science Book, Version 2.0}, 
      author={{LSST Science Collaboration}},
      year={2009},
      eprint={0912.0201},
      archivePrefix={arXiv},
      primaryClass={astro-ph.IM},
}

@article{Alonso2015BlindForeground,
  author  = {Alonso, David and others},
  title   = {Blind foreground subtraction for intensity mapping experiments},
  journal = {\mnras},
  volume  = {447},
  number  = {1},
  pages   = {400--416},
  year    = {2015},
  doi     = {10.1093/mnras/stu2474},
  archivePrefix = {arXiv},
  eprint  = {1409.8667}
}

@article{Wolz2014Foreground,
  author  = {Wolz, Laura and others},
  title   = {The effect of foreground subtraction on cosmological measurements from intensity mapping},
  journal = {\mnras},
  volume  = {441},
  number  = {4},
  pages   = {3271--3283},
  year    = {2014},
  doi     = {10.1093/mnras/stu792},
  archivePrefix = {arXiv},
  eprint  = {1310.8144}
}

@article{Irfan2021KernelPCA,
  author  = {Irfan, Melis O. and Bull, Philip},
  title   = {Cleaning foregrounds from single-dish 21 cm intensity maps with Kernel principal component analysis},
  journal = {\mnras},
  volume  = {508},
  number  = {3},
  pages   = {3551--3568},
  year    = {2021},
  doi     = {10.1093/mnras/stab2855},
  archivePrefix = {arXiv},
  eprint  = {2107.02267}
}

@article{Cunnington2023ForegroundTransfer,
  author  = {Cunnington, Steven and others},
  title   = {The foreground transfer function for H I intensity mapping signal reconstruction: MeerKLASS and precision cosmology applications},
  journal = {\mnras},
  volume  = {523},
  number  = {2},
  pages   = {2453--2477},
  year    = {2023},
  doi     = {10.1093/mnras/stad1567},
  archivePrefix = {arXiv},
  eprint  = {2302.07034}
}



\appendix

\section{Data generation parameters}
\label{app:parameters}

Table \ref{tab:survey_specs} shows the surveys specifications used to generate mock \hi\ intensity maps. In particular, we detail the frequency characteristics of the maps which are crucial for defining the redshift range and the resolution between redshift bins; the sky coverage and observational time of each survey, as well as the dishes parameters, directly impacting the \hi-to-noise ratio; and finally the temperature fluctuation range and the beam size and noise levels calculated from Equations \ref{eq:Gaussian_beam} and \ref{eq:radiometer}.

Table \ref{tab:hyper-parameters} details the fiducial values and the priors used for the cosmological and population hyper-parameters. We use a Flat Lambda CDM cosmological model and take \texttt{Planck18}~\citep{Planck2020} with BAO measurements as fiducial values. The population hyper-parameter fiducial values are motivated by the latest LVK cosmological results \citep{GWTC5_cosmo} and correspond to a Madau-Dickinson~\citep{Madau-Dickinson_2014} and "Multi-Peak" (broken power law with two Gaussian peaks)~\citep{GWTC5_cosmo} for the redshift evolution and mass models, respectively.

\begin{table}
    \centering
    \begin{tabular}{|l|c|c|c|}
    \hline
    \textbf{Survey Parameters} & \textbf{MeerKLASS-UHF} & \textbf{SKA-Mid1} \\
    \hline
    Bandwidth $[\nu_{\rm min},\nu_{\rm max}]$ [MHz]  & [580, 1000] & [350, 1050] \\
    Redshift range $[z_{\rm min},z_{\rm max}]$ & [0.420,1.449] & [0.353,3.058] \\
    Effective redshift $z_{\rm eff}$ & 0.935 & 1.706 \\
    Channel width $\delta\nu$ [MHz] & 1 & 1\\
    \hline
    Sky area $A_{\text{sky}}$ [deg$^2$] & 10,000 & 20,000 \\
    Sky fraction $f_\text{sky}$ &  0.24 & 0.48 \\
    Observation time $t_\text{obs}$ [hrs] & 2,500 & 10,000 \\
    \hline
    Number of dishes $N_\text{dish}$ &  64 & 197 \\
    Dish diameter $D_\text{dish}$ [m] &  13.5 & 15 \\
    Max scale $k_\text{min}$ [$h$ Mpc$^{-1}$] & 10$^{-3}$ & 10$^{-4}$ \\
    \hline
    $\delta T_{\rm H_I}$ range at $z_{\rm eff}$ [mK] & 0.187 & 0.119 \\
    Beam size $\sigma_\text{beam}(z_{\text{eff}})$ [deg] & 0.716 & 0.916 \\
    Noise level  $\sigma_\text{noise}(z_{\text{eff}})$ [mK] & 0.169 & 0.079
    \end{tabular}
    \caption{Specifications for the MeerKLASS-UHF and SKA-Mid1 surveys as used in the analysis. They may differ from the specifications reported in \citep{Santos_2017,SKA_Cosmology_SWG_2020}: (1) The channel width was increased to reduce the computational cost of \hi\ simulations; (2) The MeerKLASS-UHF sky area (as well as the sky fraction and observation time) was extended from 4,000 to 10,000 deg $^2$ in line with the updated observational plans \citep{Cunnington:2025sdr}; (3) The MeerKAT dishes that will form part of SKAO have $D_{\text{dish}} = 13.5$ m, but for simplicity we choose a 15 m diameter for all SKAO dishes; (4) The maximum scales were rounded. The temperature fluctuation ($\delta T_{\rm H_I}$) range is the difference between $\delta T_{\rm H_I}^{\rm min}(z_{\rm eff})$ and $\delta T_{\rm H_I}^{\rm max} (z_{\rm eff})$. The beam size and noise level values were calculated from Equations~\ref{eq:Gaussian_beam} and \ref{eq:radiometer}, respectively.}
    \label{tab:survey_specs}
\end{table}

\begin{table}
    \centering
    \begin{tabular}{l|c|c|c}
         \hline
         \textbf{Parameter} & \textbf{Notation} & \textbf{Fiducial} & \textbf{Prior}\\
         \hline
         Hubble constant & $H_0$ & 67.66 &$\mathcal{U}(10,150)$\\
         Matter density fraction & $\Omega_{\rm m}$ & 0.3111 & $\mathcal{U}(0,1)$\\
         \hline
         MD power law index $z<z_p$ & $\gamma$ & 4.6& $\mathcal{U}(0,12)$\\
         MD peak redshift & $z_p$ & 2.5& $\mathcal{U}(0,4)$\\
         MD power law index $z>z_p$ & $\kappa$ & 2.9& $\mathcal{U}(0,6)$\\
         \hline
         $m_1$ power law index  & $\alpha$ & 2 & $\mathcal{U}(1,6)$\\
         $m_2$ power law index & $\beta$ & 0.9 & $\mathcal{U}(-4,6)$\\
         Minimum mass & $m_{\rm min}$ & 5 & $\mathcal{U}(2,10)$\\
         Maximum mass & $m_{\rm max}$ & 100 & $\mathcal{U}(50,200)$\\
         Smoothing coefficient & $\delta_{\rm m}$ & 3 & $\mathcal{U}(10^{-3},10)$\\
         Mean low Gaussian peak & $\mu_{\rm g,low}$ & 10 & $\mathcal{U}(5,20)$\\
         Mean high Gaussian peak & $\mu_{\rm g,high}$ & 32 & $\mathcal{U}(20,50)$\\
         Width low Gaussian peak & $\sigma_{\rm g,low}$ & 0.5 & $\mathcal{U}(0.4,5)$\\
         Width high Gaussian peak& $\sigma_{\rm g,high}$ & 1.5 & $\mathcal{U}(0.4,10)$\\
         Gaussian peaks weight & $\lambda_{\rm g}$ & 0.4 & $\mathcal{U}(0,1)$\\
         Low Gaussian peak weight & $\lambda_{\rm g,low}$ & 0.9 & $\mathcal{U}(0,1)$\\
    \end{tabular}
    \caption{Summary of the notation, fiducial value and prior for the cosmology, Madau-Dickinson and mass model hyper-parameters. The Hubble constant has units km s$^{-1}$ Mpc$^{-1}$, the masses are in solar mass units. The primary mass, $m_1$, is set to be heavier than the secondary mass, $m_2$.}
    \label{tab:hyper-parameters}
\end{table}

\section{Foreground cleaning}
\label{app:foreground}
A major challenge for \hi\ IM is that the cosmological signal is observed together with much brighter astrophysical foreground emission. In a real survey, the measured sky temperature contains contributions from the redshifted \hi\ emission as well as galactic synchrotron and bremsstrahlung emission, extragalactic radio sources, and instrumental effects. The foreground brightness can exceed the cosmological signal by several orders of magnitude, making a direct measurement of the latter extremely challenging and foreground removal an essential preprocessing step for any subsequent analysis \citep{Wolz2014Foreground,Alonso2015BlindForeground}. Fortunately, the \hi\ signal and the foregrounds have qualitatively different behaviours across frequencies. Since frequency determines the redshift, neighbouring frequency channels probe neighbouring radial volumes and therefore contain contributions from different \hi-emitting structures, whose fluctuations are only partially correlated across frequency. The \hi\ signal consequently exhibits relatively rapid frequency fluctuations. In contrast, the dominant foreground components, such as synchrotron emission, have smooth, slowly varying spectra that can often be described approximately by power-law-like frequency dependence. This difference in spectral coherence provides a way of separating the two components without requiring a detailed physical model of the foregrounds \citep{Alonso2015BlindForeground,Irfan2021KernelPCA}. We simulate these foreground emissions and incorporate them in our simulations to eventually attempt to remove them with various levels of cleaning and assess the impact of the foreground-cleaning choices on our results.

We exploit the difference in the spectral behaviour of our two components using blind Principal Component Analysis (PCA), a statistical technique that does not assume any prior knowledge regarding the nature or characteristics of the foreground emission processes. In broad terms, PCA identifies the characteristic patterns of frequency variation that account for most of the variation in the observed data. Since the foregrounds are much brighter than the \hi\ signal and are highly correlated between neighbouring frequency channels, they dominate the variance of the data and therefore tend to be captured by the first few principal components. Removing these dominant components can consequently suppress most of the foreground contamination while retaining much of the smaller-amplitude, less spectrally coherent \hi\ signal. In our implementation, the temperature maps are arranged into a matrix $\mathbf{M}$, with frequency channels along one dimension and sky pixels along the other. After subtracting the mean frequency dependence, we calculate the empirical frequency--frequency covariance matrix, $\mathbf{C}=\mathrm{Cov}(\mathbf{M})$, and diagonalise it to obtain its eigenvectors. We construct a mixing matrix $\mathbf{A}$ from the $N_{\rm fg}$ eigenvectors with the largest eigenvalues, which span the dominant foreground subspace. Assuming orthonormal eigenvectors, the projection operator onto this subspace is $\mathbf{A}\mathbf{A}^{\mathrm{T}}$. The component of the map lying within this subspace is therefore $\mathbf{A}\mathbf{A}^{\mathrm{T}}\mathbf{M}$, which we subtract from the original map to remove the dominant foreground modes. The cleaned map is thus given by
\begin{equation}
    \mathbf{M}_{\rm clean}
    =
    \left(\mathbf{I} - \mathbf{A}\mathbf{A}^{\mathrm{T}}\right)\mathbf{M}.
\end{equation}
In this work, we consider $N_{\rm fg}=3,4,6,$ and $10$ removed modes to investigate the dependence of the results on the strength of the cleaning. The same procedure is applied to both the underlying \hi\ maps and maps including instrumental effects, allowing us to test how realistic observational effects affect the cleaning. In particular, beam smoothing and instrumental noise alter the frequency covariance from which the PCA modes are determined, and can therefore modify the separation between foreground, \hi\ signal, and noise \citep{Irfan2021KernelPCA}. 
PCA cleaning does not provide a perfect separation between foregrounds and the cosmological signal. Although the foregrounds are generally much brighter and spectrally smoother, the \hi\ signal also contains modes that vary slowly with frequency. These modes can overlap with the foreground subspace and are therefore partially removed when the dominant PCA modes are projected out. Increasing $N_{\rm fg}$ generally reduces the residual foreground contamination but also increases the amount of cosmological signal that is removed, producing a trade-off between foreground residuals and signal loss \citep{Alonso2015BlindForeground,Cunnington2023ForegroundTransfer}. This effect is particularly relevant for measurements of large-scale or slowly varying radial modes. For power-spectrum analyses, the impact of the cleaning can therefore be quantified using signal-injection simulations to determine the corresponding foreground transfer function \citep{Cunnington2023ForegroundTransfer}.


\begin{figure}
    \centering
    \includegraphics[width=\linewidth]{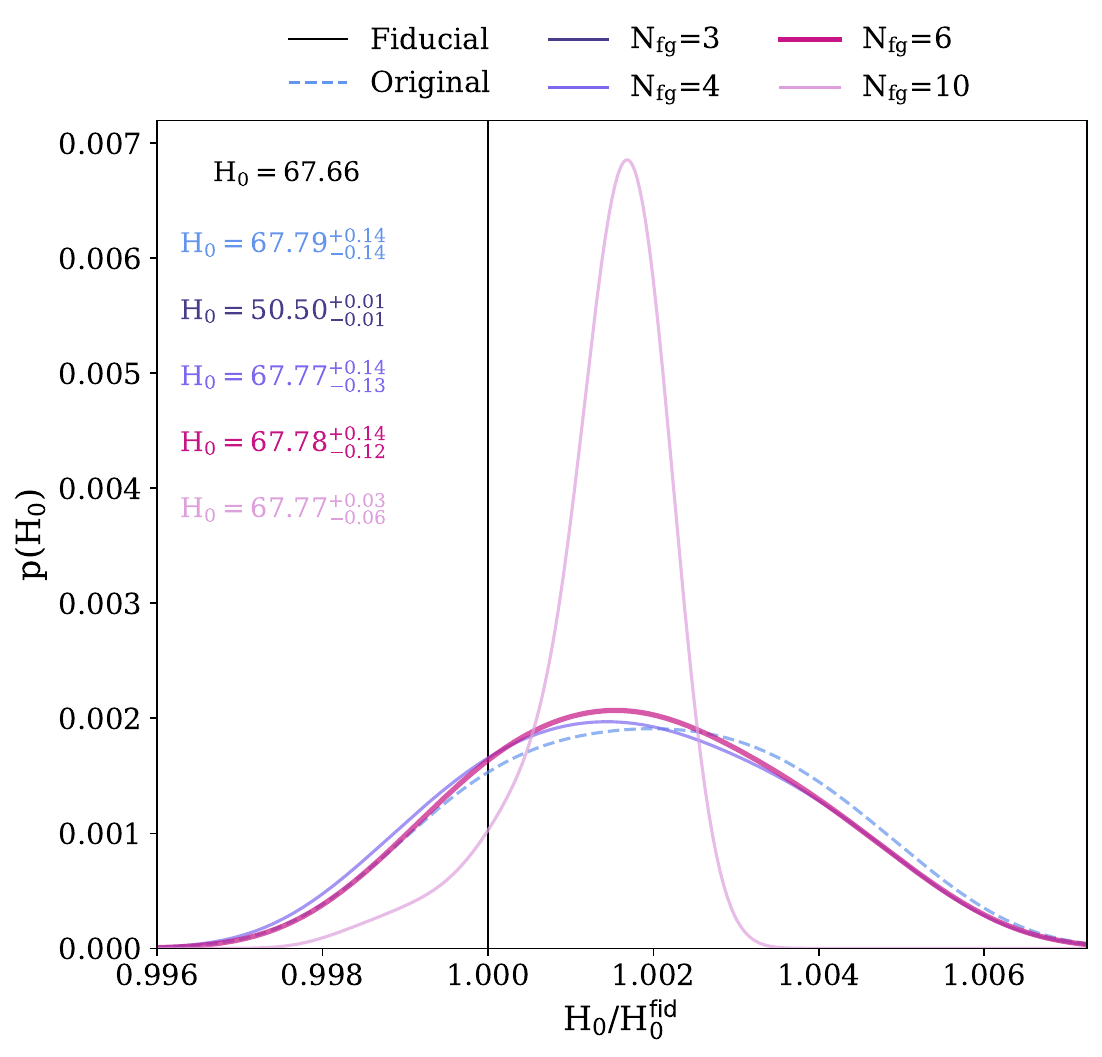}
    \caption{$H_0$ posterior distribution marginalised over all other hyper-parameters. We compare the results obtained with different values for $N_{\rm fg}$, the number of principal component used for foreground cleaning with PCA.}
    \label{fig:H0_foreground}
\end{figure}

\begin{figure}
    \centering
    \includegraphics[width=\linewidth]{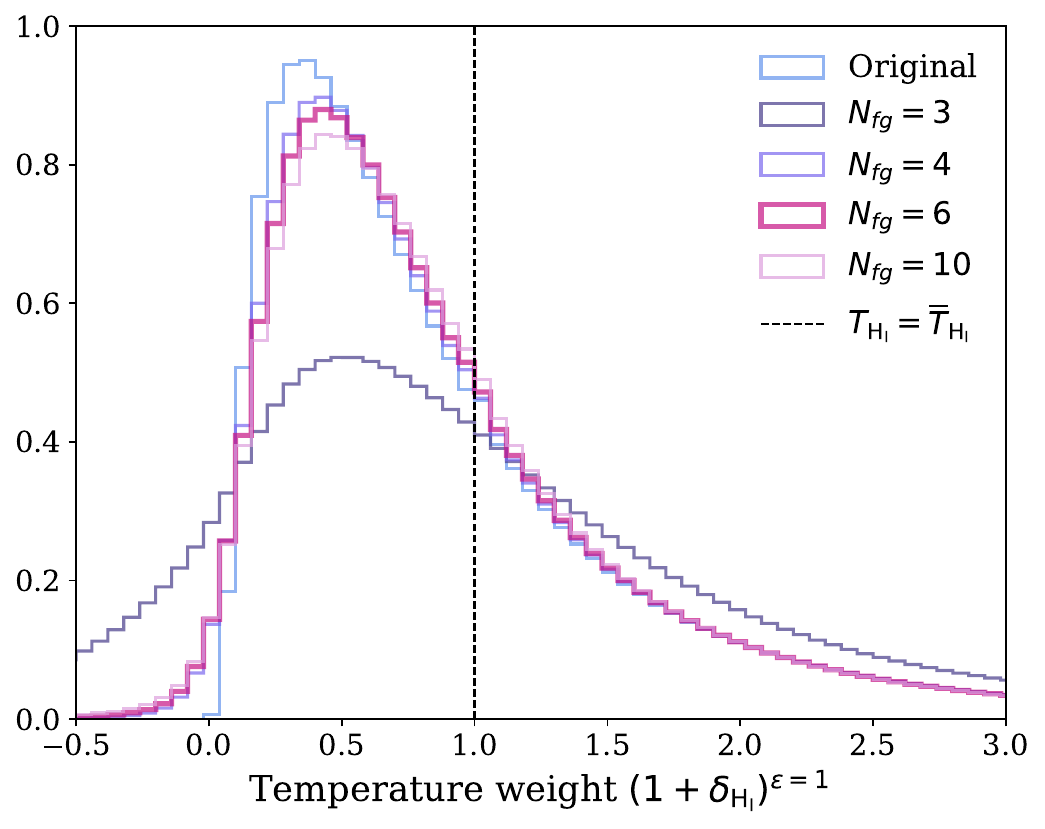}
    \caption{Temperature weight distributions for the original and foreground-cleaned SKA-Mid1 intensity maps. We compare the results obtained with different values for $N_{\rm fg}$, the number of principal component used for foreground cleaning with PCA. The dashed line shows when the density contrast equals 1 (ie. the local pixel temperature equal the mean temperature), hence, all pixels to the right (left) of the line are under-dense (over-dense).}
    \label{fig:hist_foreground}
\end{figure}

Figure~\ref{fig:H0_foreground} shows the marginalised posterior distributions for $H_0$ when the SKA-Mid1 analysis is performed with a varying level of foreground cleaning, represented by the value of $N_{\rm fg}$. A first remark is that the intermediate foreground cleaning parameters   both yield satisfactory results, with  posteriors fully compatible with the original \hi\ map case and a comparable level of precision (see Figure~\ref{fig:H0_foreground}).
In contrast, the analysis with $N_{\rm fg}=3$ is critically biased towards low values for $H_0$ while $N_{\rm fg}=10$ results in a slight bias towards high $H_0$ values. This suggests that under-cleaning or over-cleaning the foreground-contaminated map has a non-negligible impact on the cosmological inference. The foreground emission spectra are largest at low frequencies and decrease monotonically across the MeerKLASS-UHF and SKA-Mid1 frequency ranges. Since lower observing frequencies correspond to higher redshifts in 21-cm IM, foreground residuals therefore have their largest impact at high redshift. In our analysis, these residuals introduce excess structure in the intensity maps at high redshift, which can lead to an enhanced probability of high-redshift hosts in the inferred line-of-sight redshift prior. This contamination also induces significant shifts in the inferred mass-population parameters, with $m_{\rm min}$ and $\lambda_g$ being underestimated, while $m_{\rm max}$ and $\mu_g^{\rm high}$ are overestimated. These shifts are qualitatively consistent with the additional high-redshift structure introduced by the residual foregrounds, which modifies the redshift distribution of the GW sources and consequently the population parameters inferred jointly with the cosmology. This ultimately results in an underestimation of both $H_0$ and $\Omega_m$. This discrepancy with respect to the original map is illustrated by Figure~\ref{fig:hist_foreground} in which we see that the distribution of the temperature weight, used to calculate the line-of-sight prior (see Equations~\ref{eq:P(phi|z)} and \ref{eq:P(phi,z)}), displays an important departure from the initial map, in particular a wider distribution, less skewed and with long tails. Instead, the analysis with $N_{\rm fg}=10$ finds a median value for $H_0$ extremely close to that of the original \hi\ map (and actually closer to the fiducial value), only that the 68\% confidence interval is too narrow, leading to an important bias. This is particularly surprising as the effect of signal loss is not obvious in the temperature weight distribution in Figure~\ref{fig:hist_foreground}. Indeed, the distribution for $N_{\rm fg}=10$ is extremely close to $N_{\rm fg}=4 $ or $6$, with only marginally larger tails. The increase in precision seems to originate from the excessively strong degeneracy between $H_0$ and $\Omega_m$ which appears in the 2D joint posterior between the two cosmological parameters, while the population parameters are inferred similarly. If  the precision on $H_0$ increases by 68\% compared to the original case, $\Omega_m$ remains well measured with a precision augmented by 95\%. Whether signal loss is the direct cause of this curiously high precision on $\Omega_m$ is an investigation left for future work.  

For MeerKLASS-UHF, we observe a somewhat different behaviour: $N_{\rm fg}=3$ and $4$ yield results consistent with the fiducial value, whereas $N_{\rm fg}=6$ and $10$ exhibit behaviour similar to the $N_{\rm fg}=10$ case for SKA-Mid1. This suggests that there is a relatively narrow balance between under- and over-cleaning, which depends on the characteristics of the survey. The broader redshift range covered by SKA-Mid1 may participate in accurately identifying the spectral modes associated with the foregrounds, although other survey properties may also contribute to the observed differences. As in the instrumental noise case, we observe some bimodality in the posterior samples and investigate the effect of fixing $\Omega_m$, which is biased towards high values together with $H_0$, to its fiducial value. Doing so substantially reduces the bias induced by foreground cleaning, yielding a narrow peak centred on the fiducial value of $H_0$ and a secondary peak at higher $H_0$, well above the Hubble tension interval. With $N_{\rm fg}=4$, chosen as default for MeerKLASS-UHF, the fiducial peak is clearly preferred, with an amplitude approximately twice that of the secondary high-$H_0$ peak. The primary peak has a $1\sigma$ uncertainty of approximately $5\%$, indicating that the constraint around the fiducial solution becomes substantially tighter once the degeneracy with $\Omega_m$ is removed. Nevertheless, the persistence of the secondary mode highlights that foreground cleaning can significantly modify the structure of the line-of-sight prior and induce residual multimodality in the cosmological posterior.

One possible way to mitigate these mechanisms, especially in the over-cleaning case, and therefore improve the robustness over foreground removal parametrization would be to calculate $\overline{T}_{\rm H_I}$ from the cleaned map as a function of redshift, similarly to what we suggested to reduce the bias introduced by high noise maps. It is difficult to mitigate for under-cleaning considering that the foreground dominate the \hi\ signal but it is reassuring to see that only extreme under-cleaning scenarios are problematic and such cases are very easily noticeable by eye looking at the \hi\ maps. Similarly, it would be interesting to produce forecast with maps containing foreground residuals only to understand the isolate their effect on the redshift prior and cosmological inference.

\section{Impact of \hi\ Map Pixelisation} \label{app:nside}

\begin{figure}
    \centering
    \includegraphics[width=\linewidth]{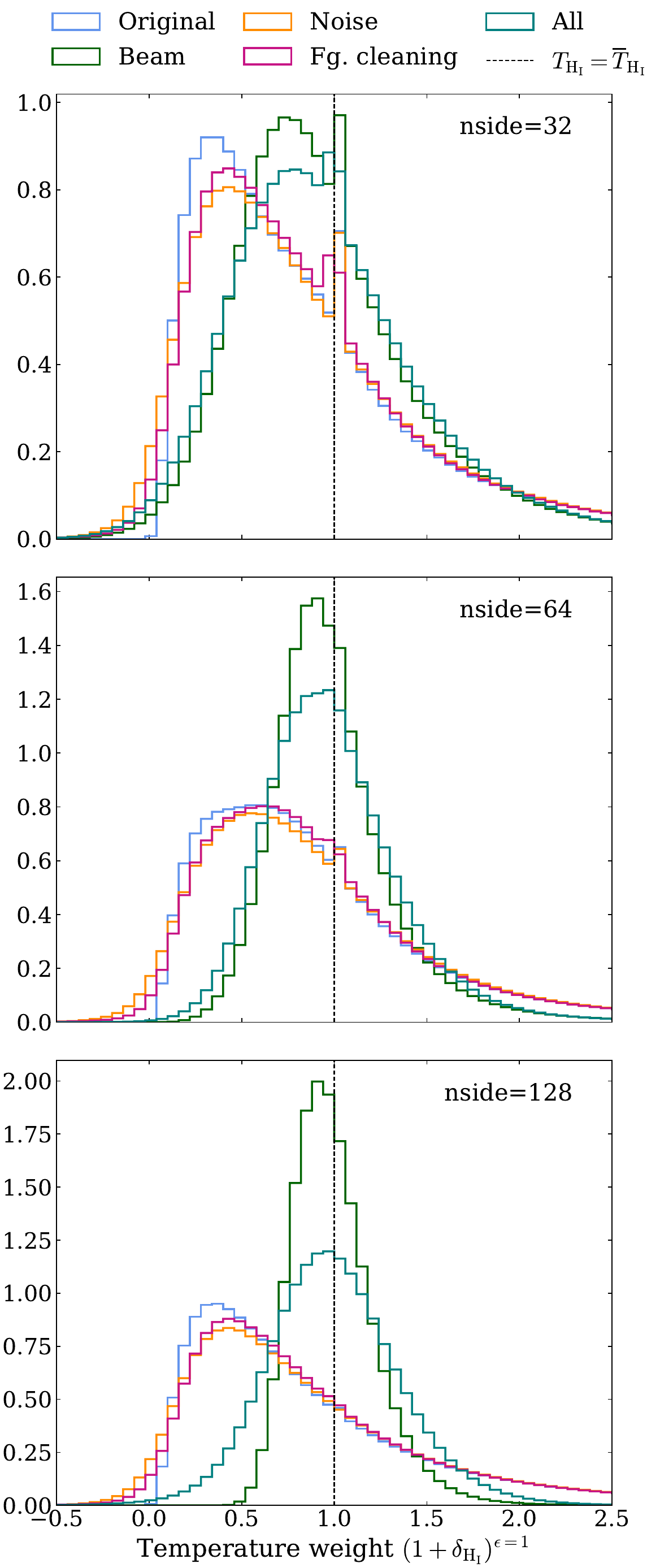}
    \caption{Impact of the \hi\ map pixelisation on the temperature-weight distribution for original \hi\ maps and maps including instrumental beam smoothing, instrumental noise, foreground cleaning, and all effects combined. We show the temperature-weight distributions, including all pixels and redshift bins, for three pixelisation resolutions, $\texttt{nside=32}$, $\texttt{64}$, and $\texttt{128}$, the resolution used in the analysis. The dashed line indicates a temperature weight of unity, corresponding to $\delta T_{\rm HI}=0$ and hence to a local pixel temperature equal to the mean temperature. Pixels to the left (right) of the line are under-dense (over-dense).}
    \label{fig:nside_study}
\end{figure}

We further investigate the dependence of the results on the map pixelisation by comparing the impact of the different observational effects for several map resolution. This comparison is performed at the map level, focusing on how the beam, instrumental noise, and foreground cleaning modify the \hi\ temperature-weight distributions as a function of the map resolution (see Figure~\ref{fig:nside_study}). The differences between the original and observed distributions are generally modest for \texttt{nside=32} and \texttt{nside=64}, while more pronounced departures are visible at \texttt{nside=128}, particularly for the highest-redshift bin. The redshift dependence is also apparent across the different map resolutions: at $z=0.35$, the distributions remain relatively similar across the different observational configurations, whereas the differences become increasingly pronounced towards $z=3.06$. This is particularly evident for the beam and combined cases, for which the high-redshift distributions can differ substantially from the corresponding original distributions. The effect of foreground cleaning also becomes more apparent at higher redshift, with changes in the shape and high-weight tail of the distribution that are less prominent at low redshift. These trends indicate that the impact of the observational effects depends on the interplay between both angular resolution and redshift.

Importantly, the map resolution considered here should not be interpreted as the intrinsic angular resolution of the \hi\ field. In a realistic \hi\ IM observation, the intrinsic \hi\ structure contains information on angular scales substantially smaller than those accessible to a single-dish survey. For example, an astrophysical object, such as a galaxy cluster, with a characteristic physical size of $10~\mathrm{Mpc}$ subtends an angle of approximately $0.32-0.37^\circ$ at the low-z limit for the MeerKLASS-UHF and SKA-Mid1 surveys, respectively, which is slightly lower than \texttt{nside=128}. The same physical scale subtends only $\sim0.09^\circ$ at $z=1.4$ and $\sim0.04^\circ$ at $z=3$, corresponding to resolutions of approximately \texttt{nside=512-1024}, respectively. This implies that the angular size of structure resolved by the pixelated maps is redshift-dependent. The effective angular resolution of the observed map is instead primarily limited by the instrumental beam, which suppresses small-scale transverse structure. For the beam sizes considered for MeerKLASS-UHF ($\sigma_{\rm beam} \in [0.515^\circ , 0.916^\circ]$) and SKA ($\sigma_{\rm beam} \in [0.458^\circ,1.374^\circ]$), the corresponding angular scales are approximately those represented by \texttt{nside=64--128}. Thus, the beam should be regarded as the physical limitation on the resolution of the observed maps, rather than the choice of \texttt{nside} itself. Therefore, the cases of \texttt{nside=64} and \texttt{nside=128} do not only correspond to lower resolution but rather to targetting larger structure scales. The results suggest that \hi\ structures with larger angular scales can be traced more reliably, by being less sensitive to observational limitations.

\bsp	
\label{lastpage}
\end{document}